# Subspace-based Identification Algorithm for Characterizing Causal Networks in Resting Brain

Shahab Kadkhodaeian Bakhtiari[a], Gholam-Ali Hossein-Zadeh[a*]

[a] Control and Intelligence Processing Center of Excellence, School of Electrical and Computer Engineering, College of Engineering, University of Tehran, Tehran 14395-515, Iran

**Corresponding author:**

Gholam-Ali Hossein-Zadeh

Address: School of Electrical and Computer Engineering,

College of Engineering, University of Tehran,

North Kargar Aven. After Jalal High Way,

Tehran, POBox: 14395-515,

Iran

Email address: ghzadeh@ut.ac.ir

Phone: (+98) -21-82084178

Fax: (+98) -21 88778690

## Abstract

The resting brain has been extensively investigated for low frequency synchrony between brain regions, namely Functional Connectivity (FC). However the other main stream of brain connectivity analysis that seeks causal interactions between brain regions, Effective Connectivity (EC), has been little explored. Inherent complexity of brain activities in resting-state, as observed in BOLD (Blood Oxygenation-Level Dependant) fluctuations, calls for exploratory methods for characterizing these causal networks. On the other hand, the inevitable effects that hemodynamic system imposes on causal inferences in fMRI data, lead us toward the methods in which causal inferences can take place in latent neuronal level, rather than observed BOLD time-series. To simultaneously satisfy these two concerns, in this paper, we introduce a novel state-space system identification approach for studying causal interactions among brain regions in the absence of explicit cognitive task. This algorithm is a geometrically inspired method for identification of stochastic systems, purely based on output observations. Using extensive simulations, three aspects of our proposed method are investigated: ability in discriminating existent interactions from non-existent ones, the effect of observation noise, and downsampling on algorithm performance. Our simulations demonstrate that Subspace-based Identification Algorithm (SIA) is sufficiently robust against above-mentioned factors, and can reliably uncover the underlying causal interactions of resting-state fMRI. Furthermore, in contrast to previously established state-space approaches in effective connectivity studies, this method is able to characterize causal networks with large number of brain regions. In addition, we utilized the proposed algorithm for identification of causal relationships underlying anti-correlation of default-mode and dorsal attention networks during the rest, using fMRI. We observed that default-mode network places in a higher order in hierarchical structure of brain functional networks compared to dorsal attention network.

# 1. Introduction

Brain's "dark energy", a concept introduced by Raichel and his colleagues in Zhang and Raichle (2010), best captures the essence of mysterious low frequency fluctuations of brain activity observed in resting-state fMRI (rsfMRI). Considering the high rate of ongoing energy consumption, compared to small increase caused by task related brain activity (Zhang and Raichle, 2010), it is expected that exploring resting brain can improve our knowledge of brain intrinsic activity.

In order to unlock the mystery of brain's "dark energy", a wide variety of machine learning and signal processing methods and algorithms have been proposed during the past 15 years. However, most of these methods are based on correlation analyses between the spontaneous oscillations of brain regions in resting-state. These methods reveal a low frequency synchrony, namely Functional Connectivity (FC) (Friston, 1994), within specific networks of brain regions. Since Biswal's seminal paper on detection of functional connectivity in motor cortex (Biswal et al., 1995), different approaches and methods have been introduced for identification of functional networks in rsfMRI. The region-of-interest (ROI) seed-based studies were the first attempts toward assessment of correlative activations in resting-state (Uddin et al., 2009; Cordes et al., 2000; Jiang et al., 2004; Biswal et al. 1997), and their simplicity and straightforward interpretation make them elegant approaches for FC studies. But the subjective selection of brain regions in these methods causes eventual biased interpretations around the a priori selected seeds. (van den Heuvel and Hulshoff Pol, 2010). Considering this crucial limitation, data-driven algorithms have been presented for estimating functional networks exempted from any requirement of prior specification of networks characteristics. Most of these methods had been successfully applied to other engineering problems, and found a good acceptability in functional connectivity analysis, as well. These methods look through functional data and autonomously find functional networks based on an informative criterion, e.g. independency between clusters of regions which form a network (Friston et al., 1993; Beckmann et al., 2005; Calhoun et al., 2001; De Luca et al., 2006; Cordes, 2002; Salvador et al., 2005). ICA, a renowned algorithm in Blind Source Separation problems, found a good acceptability in identification of independent networks in resting-state. Different networks, such as Default-Mode Network (DMN), Dorsal Attention Network (DAN), Fronto-Parietal Control Network (FPCN), etc., were detected by means of ICA algorithm as the independent sources of observed brain activity in BOLD signal (Beckmann et al., 2005).

However, the other main stream of research in brain connectivity analysis, Effective Connectivity (EC) (Friston et al., 1994), has been overlooked in resting-state studies. Effective Connectivity is introduced to represent the causal influences that each region of the brain exerts over other regions. In fact, there are some aspects of on-going brain activity which cannot be described by inadequate measures of instantaneous coupling, so causal inferences should be employed for better understanding of neuronal system. Particularly, the activation/deactivation dichotomy of brain areas which can be regularly observed in resting-state BOLD signals should be revisited in a cause and effect view within brain dynamics and structure, rather than evolved patterns of synchrony in brain activity (He and Raichle, 2009). Recent attempts in computational neuroscience for constructing bottom-up computational models of brain activity in rest have demonstrated the critical role of brain structure and internal dynamics in the emergence of observed temporal coherency (Gosh et al., 2008a, 2008b; Knock et al., 2009; Honey et al., 2009; Deco et al., 2009; Carbal et al., 2011). Even though these models suggest new ideas about resting brain based on structure-function relationship, data-driven conclusions about internal dynamics of brain activity is still required. In order to approach this viewpoint, in this paper we try to investigate dynamic behavior of resting brain in a system identification framework.

Among diverse approaches in statistical causal inference, two of them obtained more popularity in estimating EC from neuroimaging data (Valdes-sosa et al., in press): 1) Methods based on temporal precedence (Granger Causality Analysis), and 2) Methods based on physical influences (control theory and state-space approaches). A comprehensive discussion on merits and drawbacks of these methods can be found in Valdes-sosa et al. (in press), Roebroeck et al. (2009), and Friston (2009). Recent studies have endorsed the ability of state-space approaches in considering the effects of hemodynamic system on inferences about causality and the importance of this characteristic (David et al., 2008). Nevertheless, the sparse literature on EC studies in resting-state, is replete with methods based on conventional Granger-like metrics in which there is no account for the effect of hemodynamic response function (HRF) in lag information retrieval (Liao et al., 2009, 2010; Deshpande et al., 2010a). Actually, the inherent complexity of resting brain studies calls for exploratory algorithms for estimating causal dependencies in networks with large number of brain areas. Realization of this object in state-space domain in the field, has been hindered by the lack of algorithms which be able to identify high dimensional state-spaces. Established methods of state-space identification in neuroimaging studies are not suited for high dimensional state-spaces (large number of regions) (Smith et al., 2009; Ryali et al., 2010), and/or they cannot provide acceptable results in the absence of external experimental conditions (e.g. conventional DCM, Friston et al., 2003). On the other hand, considering more influential ROIs in our analysis, can reduce the rate of spurious conclusions in causality analysis caused by missing regions. Consequently, data-driven system identification algorithms are substantially needed for estimating high dimensional state-space representation of resting brain networks.

In this paper, we are going to develop a system identification method for estimating state-space representation of brain activity during rest. To the best of our knowledge, this is the first application of subspace identification methods in connectivity analysis. Despite recently established method for EC studies (Smith et al., 2009; Ryali et al., 2010), this method is not organized in an iterative regime, and there is no need for initial value selection for parameters and state variables and the drawbacks of inappropriate initialization (*e.g.,* local vs. global optimality) are avoided in this method. Moreover, the simple algebraic formulation of this algorithm makes the identification of large causal networks plausible (say 15-20 regions), and this can be helpful in decreasing the likelihood of spurious conclusions on EC detection, which can be caused by neglecting important nodes in the network. More importantly, unlike conventional state-space identification algorithms in neuroimaging, no prior information about potential causal interactions or their a priori distribution is required. This characteristic decreases the subjectivity of identification process compared to those confirmatory methods (Valdes-Sosa et al., in press).

In the following sections, we first describe our state-space model, and then we will propose the developed system identification algorithm for estimating ECs based on this model. After that, we will investigate the performance of our proposed method in detecting causal interactions in high dimensional networks through extensive simulations with different characteristics. These simulations reveal crucial specifications of this method in dealing with connectivity analysis problems. Finally, this algorithm is used for identifying causal interactions underlying observed anti-correlation between dorsal attention network and default mode network in resting brain, using fMRI data.

## 2. Materials and methods

In the following sections, firstly, we will describe the state-space model which is used throughout this paper. This model represents the time evolution of neuronal activity in absence of exogenous inputs, and transformation of this activity to observed resting-state BOLD signal. This model was proposed by Penny et. al (2005) for a univariate iterative deconvolution of neuronal activity from BOLD signal. It has also been recently used in effective connectivity analyses (Smith et al., 2009; Ryali et al., 2010). After a short review on the model, we will portray a subspace system identification framework which is developed in this study for estimation and identification of the above-mentioned state-space model.

## 2.1. State-space Model

According to recent controversies on different effective connectivity detection algorithms (Valdes-sosa et al., in press; Roebroeck et al., 2009; Friston, 2009) it seems that the main motivation in utilizing state-space representations in this domain is their ability in separate modeling of latent neuronal activity and Hemodynamic Response Function (HRF). Furthermore, this model can bring the control theory interpretation of causality to connectivity analysis, so identification of state-space models can give us an insight of causal interactions in the brain system.

The model that is used in this paper is the one proposed by Penny et al. (2005) for univariate deconvolution. In this study, we utilize multivariate version of this model (Smith et al., 2009; Ryali et al., 2010) with slight modifications that suit the model to rsfMRI. Equations (1) to (3) show the formulation of this model. The first equation represents the time evolution of neuronal dynamics in brain regions. The second equation accumulates $L$ lags of neuronal activity of $m^{th}$ region in a vector. The third equation models the convolution between neuronal time-series and HRF for generating the BOLD observations.

$$
\begin{cases}
\boldsymbol{z}[t] = \boldsymbol{A}\boldsymbol{z}[t-1] + \boldsymbol{w}[t] & (1) \\
\boldsymbol{x}_m[t] = \begin{pmatrix} \boldsymbol{z}_m[t] \\ \boldsymbol{z}_m[t-1] \\ \vdots \\ \boldsymbol{z}_m[t-L+1] \end{pmatrix} & (2) \\
y_m[t] = \boldsymbol{c}_m{}^{tr}\boldsymbol{x}_m[t] + e_m[t] & (3)
\end{cases}
$$

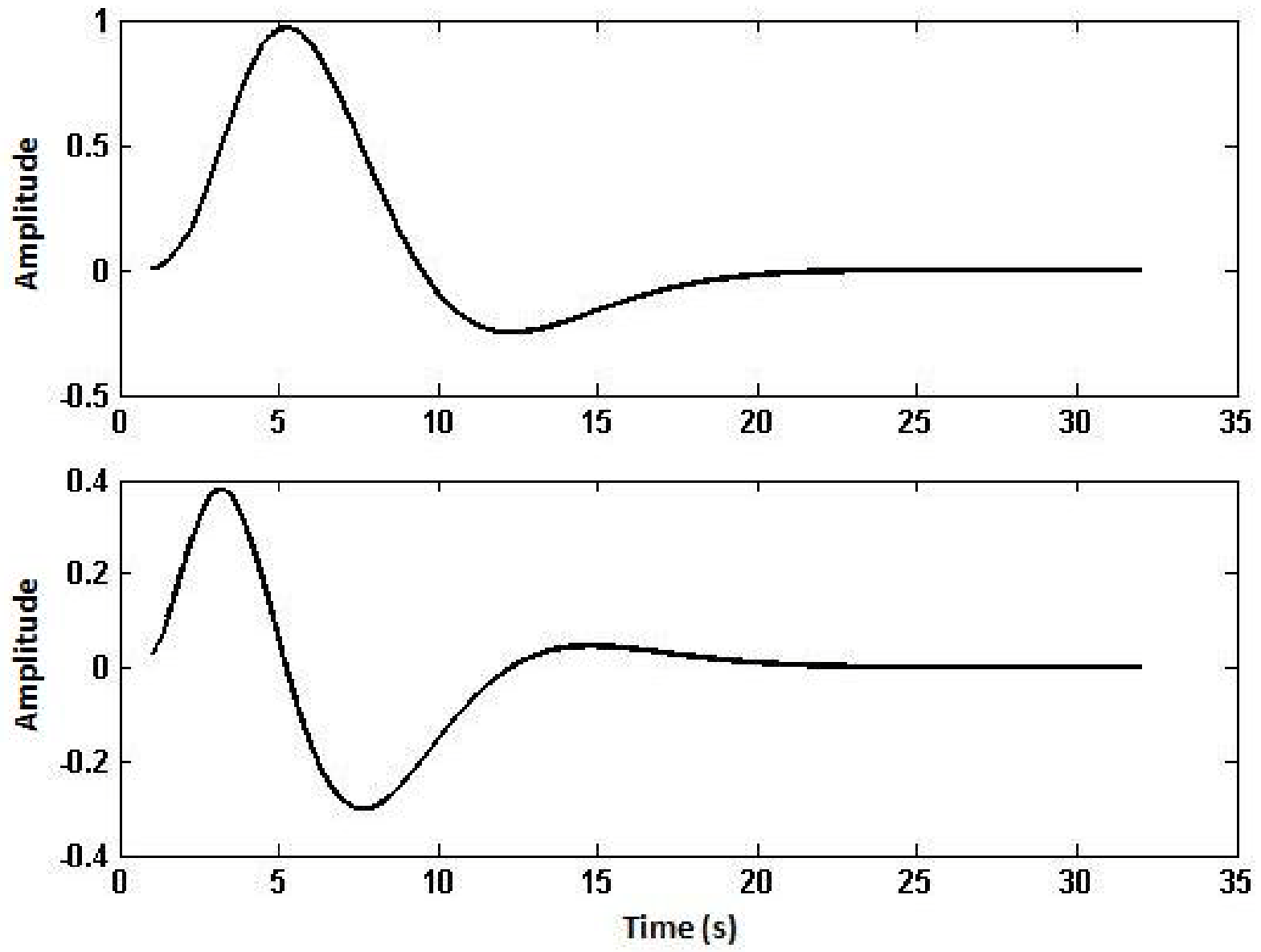

**Fig. 1.** Bases of Hemodynamic system. Top row: canonical HRF, and bottom row: time derivative of canonical HRF. Linear combination of these two vectors in our model construct HRF of each region . The variations in coefficients of this linear combination, model the inter-region variability in HRFs.

Here we represent matrices and vectors respectively by upper and lower case bold face letters. In these equations, $t$ represent discrete time instants. $\boldsymbol{z}[t]$, a $M \times 1$ vector, contains the neuronal state of $M$ regions at time $t$. Matrix $\boldsymbol{A}$ represents the strength of causal interactions between different regions, as $a_{ij}$ element of $\boldsymbol{A}$ shows the impact that region $j$ exerts over region $i$, and diagonal elements symbolize the time constant of each region's internal dynamics. The past $L$ values of $\boldsymbol{z}_m[t]$ are accumulated in $\boldsymbol{x}_m[t]$, so that the desired linear convolution between neuronal time-series and HRF will be implemented through vector inner products in equation (3). In the last equation, $y_m(t)$ is the value of BOLD signal for region $m$ at time $t$, and vector $\boldsymbol{c}_m$ contains information about HRF, and the inner product ${\boldsymbol{c}_m}^{tr}\boldsymbol{x}_m[t]$ represents the desired convolution. In other words, $\boldsymbol{c}_m$ is the hemodynamic response function for region $m$, and as the subscript $m$ explains, we presume inter-region and inter-subject variations in this vector. $\boldsymbol{w}[t]$, $M \times 1$ vector, and $e_m[t]$ are dynamic and observation noises respectively, and despite previous usages of this model, we do not assume any distribution (*e.g.*, Gaussain) for stochastic parts of the model, and we just know that they are zero mean, white vector sequences with covariance matrix:

$$\mathbf{E}\left\{\begin{pmatrix}\boldsymbol{w}[t+n]\\ e_m[t+n]\end{pmatrix}\begin{pmatrix}\boldsymbol{w}[t]\\ e_m[t]\end{pmatrix}^{tr}\right\}=\begin{pmatrix}\boldsymbol{Q} & \mathbf{0}\\ \mathbf{0} & r\end{pmatrix}\delta[n] \tag{4}$$

where $\delta[n]$ is Kronecker delta, and E is the expectation operator. As this equation shows our system identification algorithm should deal with a model characterized by more complicated dynamics, in which dynamic and observation noises can have arbitrary distribution as long as all moments are finite. These degrees of freedom in noise distribution make this model more appropriate for capturing the complexity of resting brain fluctuations.

The following formula is proposed for modeling the inter-region and inter-subject variations of HRF using two bases in figure(1):

$$\boldsymbol{c}_m = (\boldsymbol{b}_m \boldsymbol{\Phi})^{tr} \tag{5}$$

According to this equation, each region's hemodynamic response is formulated as a linear combination of convolution kernels (*e.g.,* canonical HRF and its derivatives) (Henson et al., 2001). Regarding this model, bases of hemodynamic system, rows of matrix $\boldsymbol{\Phi}$, do not vary across brain regions, and diversities in response functions are originally caused by different, $1 \times 2$ row vectors, $\boldsymbol{b}_m$. Figure (1) shows convolution kernels which have been widely used in GLM analysis of fMRI time-series for activation detection.

Since our method is developed to deal with resting-state data, compared to previous usage of this model, the effects of exogenous stimuli are neglected in equation (1), so the neuronal dynamics are modeled by means of a vector autoregressive model (VAR model), instead of a bilinear model with exogenous input.

Detecting causal interactions between $M$ regions is equivalent to estimating matrix $\boldsymbol{A}$ in this state-space model. Each element of this matrix, except diagonal ones, explains the strength of effective connectivity between related regions. Therefore, retrieving the magnitude of matrix $\boldsymbol{A}$ elements from BOLD observations can give an insight of the causal interactions within neuronal system. Furthermore, state-space representation of BOLD time-series, using equations (1)-(3), can ensure the consideration of hemodynamic system influences on causality inference. These influences are mostly caused by low-pass filtering effect of hemodynamic system which consequently can lead to spurious inferences about causal interactions (Deshpande et al., 2010b). In order to decrease the hemodynamic system implications on causality inference, state-space approaches have been successful in simultaneous estimation of directional connectivity and HRF parameters of each region. This simultaneous estimation considers the possible dependencies between dynamics in neuronal level and variability in HRFs, and gives us more accurate results (David et al., 2008).

## 2.2. Embedded Linear State-space model

The state-space model presented in equations (1)-(3) can be reformulated in standard state-space representation (Kalman, 1960) based on embedded state variable $\boldsymbol{x}[t]$. This standard formulation is necessary for implementation of linear state-space system identification algorithms. Following equations describe this representation:

$$\begin{cases} \boldsymbol{x}[t] = \widetilde{\boldsymbol{A}}\boldsymbol{x}[t-1] + \widetilde{\boldsymbol{w}}[t] & (6) \\ \boldsymbol{y}[t] = \boldsymbol{C}\boldsymbol{x}[t] + \boldsymbol{e}[t] & (7) \end{cases}$$

In this equation, $\boldsymbol{x}[t]$, $\widetilde{\boldsymbol{A}}$, $\widetilde{\boldsymbol{w}}[t]$, $\boldsymbol{y}[t]$, $\boldsymbol{C}$, and $\boldsymbol{e}[t]$ are constructed as follows:

$$\boldsymbol{x}[t] = \begin{pmatrix} \boldsymbol{z}_1[t] \\ \boldsymbol{z}_2[t] \\ \vdots \\ \boldsymbol{z}_M[t] \\ \boldsymbol{z}_1[t-1] \\ \boldsymbol{z}_2[t-1] \\ \vdots \\ \boldsymbol{z}_M[t-1] \\ \vdots \\ \boldsymbol{z}_1[t-L+1] \\ \boldsymbol{z}_2[t-L+1] \\ \vdots \\ \boldsymbol{z}_M[t-L+1] \end{pmatrix} \quad (8)$$

$$\widetilde{\boldsymbol{A}} = \begin{pmatrix} \boldsymbol{A} & \boldsymbol{0}_{M \times M(L-1)} \\ \boldsymbol{I}_{M(L-1)} & \boldsymbol{0}_{M(L-1) \times M} \end{pmatrix} \quad (9)$$

$$\widetilde{\boldsymbol{w}}[t] = \begin{pmatrix} \boldsymbol{w}[t] \\ \boldsymbol{0}_{M(L-1) \times 1} \end{pmatrix} \quad (10)$$

$$\boldsymbol{e}[t] = \begin{pmatrix} e_1[t] \\ \vdots \\ e_M[t] \end{pmatrix} \quad (11)$$

$$\boldsymbol{y}[t] = \begin{pmatrix} y_1[t] \\ \vdots \\ y_M[t] \end{pmatrix} \quad (12)$$

$$\boldsymbol{C} = \boldsymbol{B}\widetilde{\boldsymbol{\Phi}} \quad (13)$$

In the equations above, $\boldsymbol{I}_n$ denotes $n \times n$ identity matrix, and $\boldsymbol{B}$ and $\widetilde{\boldsymbol{\Phi}}$ are defined as follows:

$$\boldsymbol{B} = \begin{pmatrix} \boldsymbol{b}_1 & \cdots & \boldsymbol{0} \\ \vdots & \ddots & \vdots \\ \boldsymbol{0} & \cdots & \boldsymbol{b}_M \end{pmatrix} \tag{14}$$

$$\widetilde{\boldsymbol{\Phi}} = \begin{pmatrix} \varphi_{11} & \boldsymbol{0} & \varphi_{12} & \boldsymbol{0} & \cdots & \varphi_{1L} & \boldsymbol{0} \\ \varphi_{21} & \boldsymbol{0} & \varphi_{22} & \boldsymbol{0} & \cdots & \boldsymbol{0} & \varphi_{2L} \\ \boldsymbol{0} & \varphi_{11} & \boldsymbol{0} & \varphi_{12} & \cdots & \varphi_{1L} & \boldsymbol{0} \\ \boldsymbol{0} & \varphi_{21} & \boldsymbol{0} & \varphi_{22} & \cdots & \boldsymbol{0} & \varphi_{2L} \\ \vdots & \vdots & \vdots & \vdots & & \vdots & \vdots \end{pmatrix} \tag{15}$$

Based on this standard structure of state-space, we can treat this system as a linear state-space model, and any algorithm developed for linear state-space models can be applied to this system, too. In the next section, we will present Subspace-based Identification Algorithm (SIA) for identifying linear state-space models. Using this system identification algorithm, we estimate matrix $\widetilde{\boldsymbol{A}}$, and implicitly $\boldsymbol{A}$, as quantitative representations of effective connectivity between resting-state BOLD time-series.

## 2.3. Subspace method for state-space system identification

The algorithm developed here is a geometrically inspired method for identification of stochastic systems, purely based on output observations $\{\boldsymbol{y}[t]\}_{t=1}^{T}$. Where system matrices $\{\widetilde{\boldsymbol{A}}, \boldsymbol{C}\}$ in (6)-(7) are known, conventional iterative Kalman filtering can retrieve state vector sequence (Kalman, 1960). However, in the context of subspace method, it is proved that in linear state estimation these system matrices are not required, and mere output observations contain sufficient information for estimating latent states (van Overschee and De Moor, 1996). A major advantage of this algorithm is the non-iterative formula for direct state estimation from rsfMRI observations. Despite iterative Kalman filtering, setting initial values for parameters and states is not a concern anymore, and consequently, drawbacks of inappropriate initialization will not affect our final inferences.

Before describing the formulation, we need to introduce some notations that will be used in the procedure.

### 2.3.1. Definitions

*Block Hankel Matrices*

Block Hankel matrices play an important role in SIA. $\boldsymbol{Y}_{0|2i-1}$ denotes output block Hankel matrix, and can be constructed from vertical augmentation of $\boldsymbol{Y}_p$ and $\boldsymbol{Y}_f$ using purely output observations as follows:

$$\boldsymbol{Y}_p = \begin{pmatrix} \boldsymbol{y}[1] & \dots & \boldsymbol{y}[j] \\ \vdots & \ddots & \vdots \\ \boldsymbol{y}[i] & \dots & \boldsymbol{y}[i+j] \end{pmatrix} \tag{16}$$

$$\boldsymbol{Y}_f = \begin{pmatrix} \boldsymbol{y}[i+1] & \dots & \boldsymbol{y}[i+j+1] \\ \vdots & \ddots & \vdots \\ \boldsymbol{y}[2i] & \dots & \boldsymbol{y}[2i+j+1] \end{pmatrix} \tag{17}$$

$$\boldsymbol{Y}_{0|2i-1} = \begin{pmatrix} \boldsymbol{Y}_p \\ -- \\ \boldsymbol{Y}_f \end{pmatrix} \tag{18}$$

Also, we need to define the matrices in the structure below:

$$\boldsymbol{Y}_p^+ = \begin{pmatrix} \boldsymbol{y}[1] & \dots & \boldsymbol{y}[j] \\ \vdots & \ddots & \vdots \\ \boldsymbol{y}[i+1] & \dots & \boldsymbol{y}[i+j+1] \end{pmatrix} \tag{19}$$

$$\boldsymbol{Y}_f^- = \begin{pmatrix} \boldsymbol{y}[i+2] & \dots & \boldsymbol{y}[i+j+2] \\ \vdots & \ddots & \vdots \\ \boldsymbol{y}[2i] & \dots & \boldsymbol{y}[2i+j+1] \end{pmatrix} \tag{20}$$

In these equations $i$ and $j$ are user-defined indices, and their value depend on the size of available observations. The value of $i$ should be chosen at least bigger than the number of brain regions in our model, and $j$ is typically equal to $T-2i$ which guaranties the usage of all observation data points. It is noteworthy that the bigger we choose $i$, the more reliable estimations we can obtain for identified system (De Moor, 2003).

The subscript "p" stands for "past", and the subscript "f" for "future". Despite this notation, there isn't any firm boundary between past and future output values in (16)-(17) and (19)-(20), and past and future outputs have many elements in common. However, there isn't any element in common between corresponding columns of $\boldsymbol{Y}_p$ and $\boldsymbol{Y}_f$, or $\boldsymbol{Y}_p^+$ and $\boldsymbol{Y}_f^-$.

*Observability Matrix*

The observability matrix, which is defined in control theory for determining identifiable modes of a state-space representation, is also heavily used in subspace algorithm, and can be constructed as in the below equation:

$$\boldsymbol{\Gamma}_i = \begin{pmatrix} \boldsymbol{C} \\ \boldsymbol{C}\widetilde{\boldsymbol{A}} \\ \boldsymbol{C}\widetilde{\boldsymbol{A}}^2 \\ \vdots \\ \boldsymbol{C}\widetilde{\boldsymbol{A}}^{i-1} \end{pmatrix} \tag{21}$$

*Kalman filter state sequence*

This state sequence can be estimated by the SIA, and is equal to estimated state sequence by a set of non-steady state Kalman filters working in parallel on each of the columns of block Hankel matrix of past outputs $\boldsymbol{Y}_p$. This state sequence is denoted by equation below:

$$\widehat{\boldsymbol{X}}_i = (\widehat{\boldsymbol{x}}[i] \quad \widehat{\boldsymbol{x}}[i+1] \quad \dots \quad \widehat{\boldsymbol{x}}[i+j-1]) \tag{22}$$

### 2.3.2. Subspace-based Identification Algorithm (SIA)

Based on these definitions, now we can develop the formulation for Subspace-based Identification Algorithm. Detailed information on these algorithms and discussions on differences between various implementations of subspace methods can be found in (van Overschee and De Moor, 1996; Katayama, 2005; Moonen et al., 1989; De Moor, 2003; Moonen, 1990).

Equations (23) and (24) calculate the projection of future block Hankel matrix of output on its past Hankel matrix:

$$\boldsymbol{P}_i = \left(\boldsymbol{Y}_f {\boldsymbol{Y}_p}^{tr}\right)\left(\boldsymbol{Y}_f {\boldsymbol{Y}_p}^{tr}\right)^{\dagger} \boldsymbol{Y}_p \tag{23}$$

$$\boldsymbol{P}_{i-1} = \left(\boldsymbol{Y}_f^- {\boldsymbol{Y}_p^+}^{tr}\right)\left(\boldsymbol{Y}_f^- {\boldsymbol{Y}_p^+}^{tr}\right)^{\dagger} \boldsymbol{Y}_p^+ \tag{24}$$

where $(J)^{\dagger}$ denotes the Moore-Penrose pseudo-inverse of matrix $J$.

After Singular Value Decomposition (SVD) on $\boldsymbol{P}_i$:

$$\boldsymbol{W}_1 \boldsymbol{P}_i \boldsymbol{W}_2 = \boldsymbol{U}_P \boldsymbol{S}_P \boldsymbol{V}_P \tag{25}$$

We choose first $ML$ biggest singular values in matrix $\boldsymbol{S}_P$, and their corresponding columns in $\boldsymbol{U}_P$. $M$ denotes the number of brain regions in our model, $L$ denotes the time length of HRF, and $\boldsymbol{U}_P^M$ and $\boldsymbol{S}_P^M$ respectively represent first $ML$ columns of $\boldsymbol{U}_P$ and matrix of first $ML$ singular values**.**

In (25) weighting matrices $\boldsymbol{W}_1$ and $\boldsymbol{W}_2$ have crucial effects on the estimation of vector sequence and observabilty matrix. There are intricate discussions in the literature on the role of the weighting matrices in identified state-space that we will not cover them here, but it should be mentioned that they determine the state-space basis in which the model will be identified (van Overschee and De Moor, 1996). In this application we will consider following values for weighting matrices:

$$\boldsymbol{W}_1 = \boldsymbol{I}_{MLi} \tag{26}$$

$$\boldsymbol{W}_2 = \boldsymbol{I}_j \tag{27}$$

Now we can construct observability matrix using equation below:

$$\boldsymbol{\Gamma}_i = \boldsymbol{W}_1{}^{-1}\boldsymbol{U}_P^M(\boldsymbol{S}_P^M)^{\frac{1}{2}}. \tag{28}$$

and stripping the last $ML$ rows of $\boldsymbol{\Gamma}_i$, we can find $\boldsymbol{\Gamma}_{i-1}$.

After calculating observability matrix $\boldsymbol{\Gamma}_i$, and orthogonal projection $\boldsymbol{P}_i$, we can determine state vector sequence through following equation:

$$\widehat{\boldsymbol{X}}_i = \boldsymbol{\Gamma}_i{}^{\dagger}\boldsymbol{P}_i \tag{29}$$

$$\widehat{\boldsymbol{X}}_{i+1} = \boldsymbol{\Gamma}_{i-1}{}^{\dagger}\boldsymbol{P}_i \tag{30}$$

Given estimated state vector sequence, we can obtain system matrices $\{\widetilde{\boldsymbol{A}}, \boldsymbol{C}\}$:

$$\begin{pmatrix} \widetilde{\boldsymbol{A}} \\ \boldsymbol{C} \end{pmatrix} = \begin{pmatrix} \widehat{\boldsymbol{X}}_{i+1} \\ \boldsymbol{Y}_{i|i} \end{pmatrix} . \widehat{\boldsymbol{X}}_i^{\dagger} \tag{31}$$

The equations (16)-(31) sketch out Subspace-based Identification Algorithm.

Here we should stress the fact that despite the impulse response of linear systems, state-space description is not unique. Therefore the estimated state vector sequences and system matrices, $\{\widetilde{\boldsymbol{A}}, \boldsymbol{C}\}$ in above procedure, might be equal to original ones up to a similarity transformation (van Overschee and De Moor, 1996). To be more specific, we assume that the state transformation $\boldsymbol{x}[t] \mapsto \boldsymbol{T}\boldsymbol{x}[t]$ can change the main state-space representation to the realization below:

$$\begin{cases} \boldsymbol{x}[t] = \boldsymbol{T}^{-1}\widetilde{\boldsymbol{A}}\boldsymbol{T}\boldsymbol{x}[t-1] + \widetilde{\boldsymbol{w}}'[t] & (32) \\ \boldsymbol{y}[t] = \boldsymbol{C}\boldsymbol{T}\boldsymbol{x}[t] + \boldsymbol{e}[t] & (33) \end{cases}$$

where $\boldsymbol{T}$ is an invertible matrix. It is not hard to show that all the systems described by (32)-(33), with various non-singular transform matrices, have same impulse response or input/output transfer function. Actually, the developed subspace method is able to identify our state-space model in the structure of above general realization (van Overschee and De Moor, 1996), and finding the best transformation matrix $\boldsymbol{T}$ which can transform our identified state-space to desired realization, presented through equations (6)-(15) is the main challenge in this context. We present identified system with below equations:

$$\begin{cases} \widehat{\boldsymbol{x}}[t] = \widehat{\boldsymbol{A}}\widehat{\boldsymbol{x}}[t-1] + \widetilde{\boldsymbol{w}}[t] & (34) \\ \boldsymbol{y}[t] = \widehat{\boldsymbol{C}}\widehat{\boldsymbol{x}}[t] + \boldsymbol{e}[t] & (35) \end{cases}$$

Given this estimated state-space, we look for the transformation matrix that transform the above state-space representation to our desired realization with specifications described through equations (6)-(15). Putting $\hat{\boldsymbol{x}}[t] = \boldsymbol{T}\boldsymbol{x}[t]$ in (34) and (35), following equations can be obtained that describe the relationship between estimated system matrices $\{\widehat{\boldsymbol{A}}, \widehat{\boldsymbol{C}}\}$ and system matrices with desired structure $\{\widetilde{\boldsymbol{A}}, \boldsymbol{C}\}$:

$$\begin{cases} \boldsymbol{x}[t] = \boldsymbol{T}^{-1}\widehat{\boldsymbol{A}}\boldsymbol{T}\boldsymbol{x}[t-1] + \widetilde{\boldsymbol{w}}'[t] & (36) \\ \boldsymbol{y}[t] = \widehat{\boldsymbol{C}}\boldsymbol{T}\boldsymbol{x}[t] + \boldsymbol{e}[t] & (37) \end{cases}$$

According to these equations and equations (6)-(7), it is obvious that $\boldsymbol{T}^{-1}\widehat{\boldsymbol{A}}\boldsymbol{T} \equiv \widetilde{\boldsymbol{A}}$, and $\widehat{\boldsymbol{C}}\boldsymbol{T} \equiv \boldsymbol{C}$. Based on this formulas and the structures we described for $\widetilde{\boldsymbol{A}}$ and $\boldsymbol{C}$ in (9) and (13), we have developed a numerical optimization-based algorithm for finding the best linear transformation matrix $\boldsymbol{T}$. Detailed equations and steps of this algorithm can be found in Appendix A.

## 2.4. Statistical Inference

To determine the significance of estimated elements of matrix $\boldsymbol{A}$ that represent effective connectivity between brain regions, we define a hypothesis testing problem, in which the null hypothesis ($H_0$) indicates the insignificance of element $a_{ij}$ ($a_{ij} = 0$). Surrogate data is generated for estimating the distribution of $a_{ij}$ under the null hypothesis. We generated surrogate data by phase shuffling the BOLD time-series. This preserves the correlations and variance of the original data but destroys any statistical dependencies that would be mediated by effective connectivity among regions.

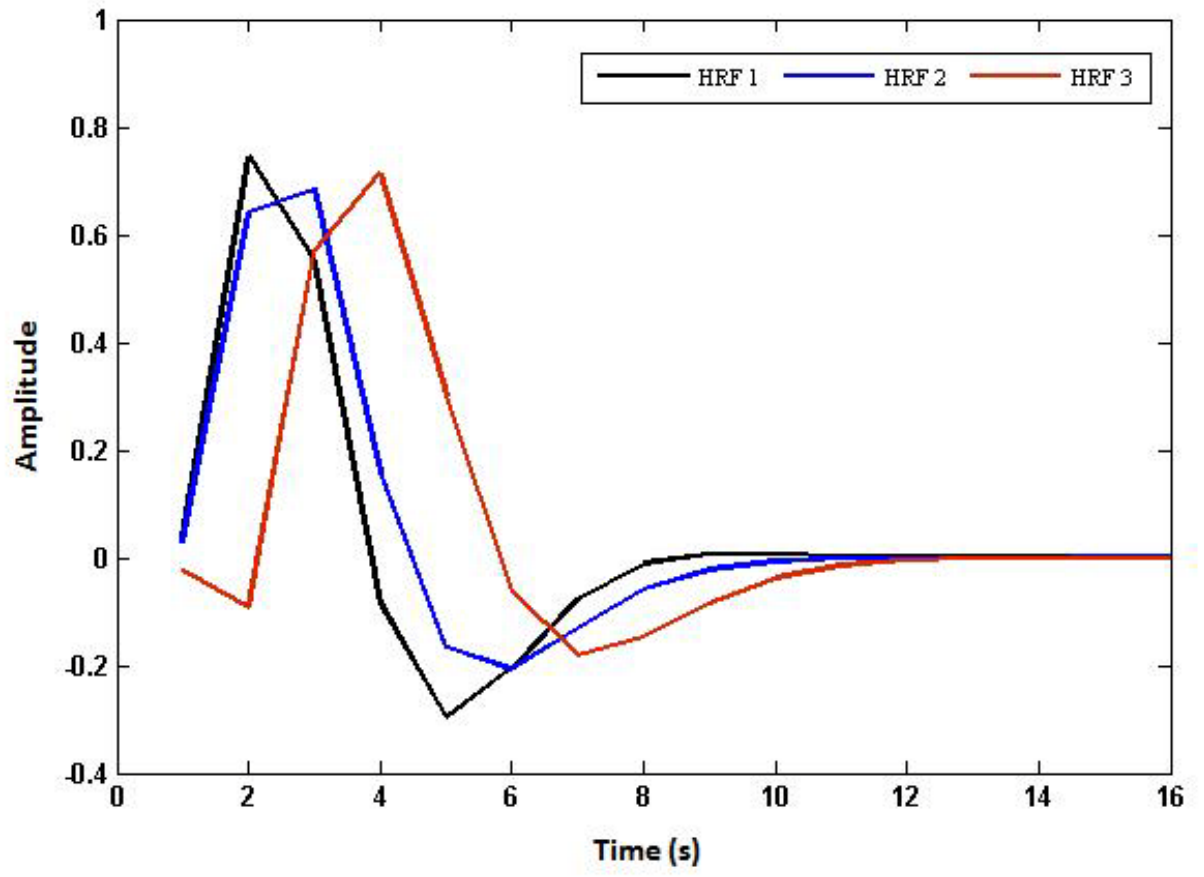

**Fig. 2.** Simulated HRFs. Above displayed HRFs are linear combinations of hemodynamic system basis (Canonical HRF and its time derivative) used in simulations. Three different values for vector $\boldsymbol{b}_m$ have been used for modeling above HRFs.

Significance of each element is determined through below equation (Theiler et al., 1992):

$$\boldsymbol{S} = |\boldsymbol{A} - \boldsymbol{\Omega}|./\boldsymbol{\Sigma} \tag{38}$$

where $\boldsymbol{A}$ is the estimated connectivity matrix, $\boldsymbol{\Omega}$ is the sample mean of surrogate values of connectivity matrix, $\boldsymbol{\Sigma}$ is the sample standard deviation of connectivity matrix surrogate values, and "./" denotes the element by element matrix division. Furthermore, error bars ($\Delta\boldsymbol{S}$) can be calculated for the values of significance. They can be very informative in determining ECs, particularly for those close to threshold value:

$$\Delta\boldsymbol{S} = \sqrt{(1 + \frac{1}{2}\boldsymbol{S}^{\boldsymbol{2}})/N} \tag{39}$$

In the above equation, $N$ represents the number of surrogate data realizations (in this study $N = 500$). Based on this equations, for estimated matrix $\boldsymbol{A}$, firstly, we calculated matrix $\boldsymbol{S}$, that represents the significance of each element in matrix $\boldsymbol{A}$. Based on p-values, we then set a threshold for determining significant ECs (*e.g.*, $\alpha < 0.01$). For ECs which are above the threshold, we also check error bars, so that we can be assured that they are significantly above the threshold. It is worth noting that finite number of surrogate data ($N$) makes us consider uncertainty in form of error bars in determining significant connections. It is obvious from (39) that for infinite number of surrogate data ($N \to \infty$) error bars will approach zero. It should be mentioned that for a network with $M$ regions, surrogate data is generated by replacing the phase of Fourier transform of time-series at each frequency $f$ by an independent random variable $\varphi$ with uniform distribution in the range $[0, 2\pi)$. As explained above, this process leads to elimination of existent interactions between original time-series, while the characteristics of power spectrum are still preserved.

## 2.5. Simulation Dataset

To evaluate the performance of SIA, we simulate the BOLD outputs of some example causal networks. To do so, we used equations (1)-(3) for modeling the dynamics of neuronal and output hemodynamic system.

The most important factor, which has been the main motivation for proposing the developed method, is the size of network and its potential effects on the accuracy of EC detection. In order to measure the degradation that occurs in the accuracy of subspace method due to increasing the number of regions,, we simulated networks with $M = 5, M = 10, M = 12,$ and $M = 15$ regions. BOLD time-series is simulated via equations (1)-(3). Input is i.i.d zero mean multivariate Gaussian noise. Variability in HRFs among different regions is considered through assigning different values to matrix $\boldsymbol{b}_m$ for different regions. The topologies of causal simulated networks are determined randomly constrained by the stability of simulated network (eigenvalues of VAR model must be inside the unit circle) . Therefore, the off-diagonal elements of matrix $\boldsymbol{A}$ take any value except those that violate the stability constraint of neuronal dynamics. Below is a typical simulated matrix $\boldsymbol{A}$ for a network with five nodes:

$$\boldsymbol{A} = \begin{pmatrix} 0.7 & 0.31 & 0 & 0 & -0.46 \\ 0 & 0.7 & 0 & 0 & 0 \\ 0 & 0 & 0.7 & 0.55 & 0 \\ -0.38 & 0 & 0 & 0.7 & 0 \\ 0 & 0 & 0 & 0 & 0.7 \end{pmatrix}$$

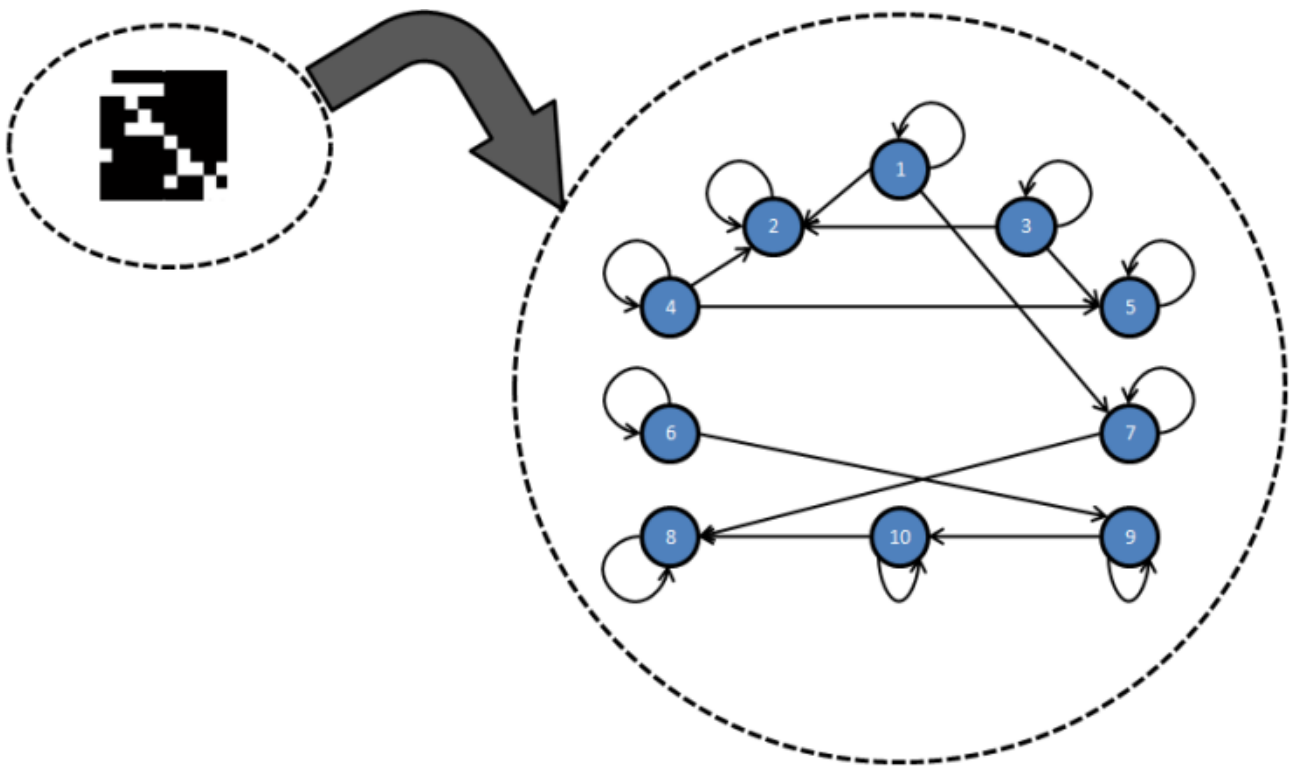

**Fig. 3.** Sample topology network used in one of the simulations. The black and white square displays the connectivity matrix. White and black squares respectively show non-zero and zero elements in matrix $\boldsymbol{A}$. The schematic connectivity pattern corresponding to this connectivity matrix is also depicted in this figure.

For each number of regions $M$, we simulated BOLD signal for 50 times. These 50 simulations (simulated subjects) differ in stochastic inputs and inter-subject variability in hemodynamic responses. This variability is introduced in HRFs by choosing the elements of matrix $\boldsymbol{b}_m$ randomly from a uniform distribution with standard deviation of 0.1 and specific mean for each region. With the purpose of decreasing biased conclusions toward a specific topology, for each number of regions $M$, we generated simulated BOLD time-series for different random topologies. Thus the variation in SIA performance for different levels of network complexity is partially considered. We have summarized the specifications of simulated dataset in Table (1). In figure (2) three samples of HRFs which have been used in above-mentioned simulations are displayed. Figure (3) indicates a sample topology which has been used in one of the simulations. The length of the simulated time-series used in this study is $T = 300\ s$.

Another paramount factor which should be taken into consideration in these simulations is the effect of the power of the observation noise on EC detection accuracy. To investigate this effect, we maintained all other factors for each network topology, and produced several set of BOLD time-series with different levels of observed Signal to Noise Ratio (SNR):

$$SNR = 20 * \log_{10} \frac{Y_m}{\sigma_m} \tag{40}$$

where $Y_m$ and $\sigma_m$ respectively denote the maximum level of BOLD time-series, and the standard deviation of observation noise in $m^{th}$ region. For each SNR level, we simulated 50 datasets which their difference is just in the pattern of random generated observation noise. We try to demonstrate how decreases in SNR level can affect the SIA performance in EC detection. In the next section we introduce the metrics that can reveal this relationship based on SIA's efficiency in dealing with these simulated datasets.

Given the fact that BOLD fMRI data is generated by downsampling the neuronal activity after convolving with region specific HRFs, we also examined the effect of downsampling on the accuracy of EC detection. To this end we adopted the approach described by Deshpandeh et al. (2009). We generated 50 simulated neuronal fluctuations at 1 KHz sampling rate for a specific network topology, shown in figure (4). Then, we convolved the simulated signals with region specific HRFs at the same sampling rate, and then downsampled for generating BOLD time-series using three different sampling rates (TR=1, 2, 3 seconds). In the identification process, these different sampling rates correspond to different embedding dimensions $L$ which are also the lengths of HRFs, and can be obtained by downsampled versions of canonical HRF.

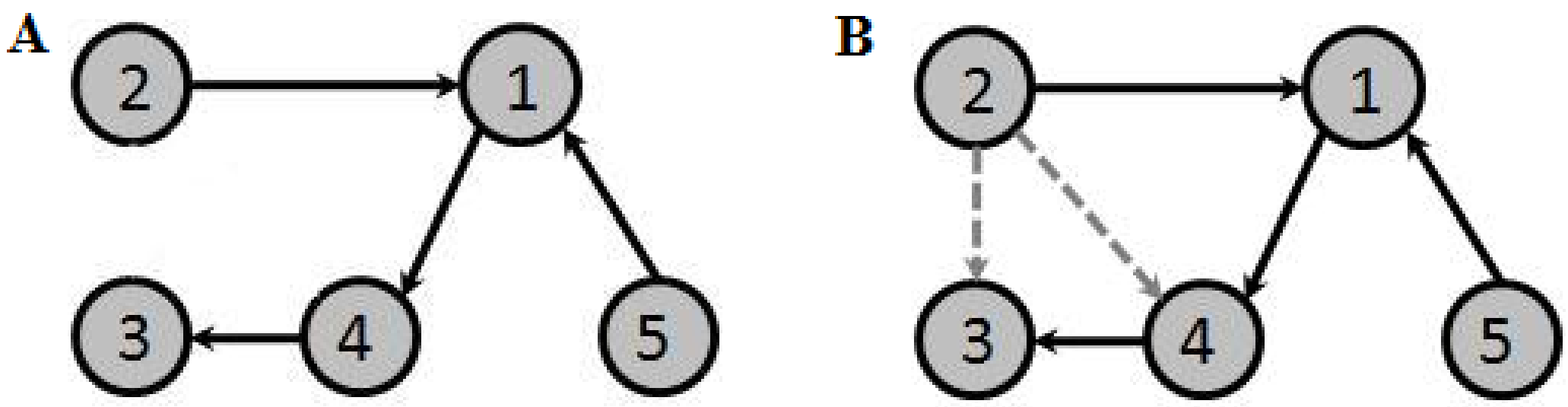

**Fig. 4. (A)** This topology is used for two purposes in this study. First, we used this topology for investigating the effect of TR on SIA performance. BOLD signal for all the regions in this network is simulated for $TR = 1, 2, 3\ seconds$, and the accuracy of SIA in the detection of effective connectivity in this network is displayed in figure (8). Second, we used this network for examining the effect of temporal aggregation caused by HRF convolution on SIA and cGCA results. **(B)** In this figure the dotted connections are prone to be falsely detected as existent connections by conditional Granger Causality Analysis (cGCA) while increasing the latency of HRF in region 2

## 2.6. Experimental Resting-State fMRI Data

The rsfMRI data that we used in this study is being made freely available through ADHD-200 project and The Neuro Bureau from the International Neuroimaging Data-sharing Initiative and are licensed with Attribution Non-commercial Creative Commons License. We selected 61 healthy subjects (28 males and 33 females, with age 9.4 $\pm$ 0.4 years) from Kennedy Krieger Institute (KKI) dataset. The scans lasted either 5 minutes and 20 seconds or 6 minutes and 30 seconds. Participants were instructed to relax, stay as still as possible, keep eyes open, and fixate on a center cross. A $T_2^*$-weighted echo planar imaging sequence was used with the following parameters: matrix of $84 \times 81$, 47 axial slices without gap and with thickness of 3 mm, flip angle $75°$, FOV of $256$, TR/TE of 2500 ms/30 ms. Preprocessing steps including slice-timing correction, motion correction, band-pass filtering of time series ($0.009\ Hz < f < 0.08\ Hz$), 6 mm full width half max Gaussian smoothing, and regressing out the WM and CSF time courses from time-series using WM and CSF masks were carried out on the data. Time courses of ROIs were extracted from preprocessed resting-state data using Automatic Anatomical Labeling (AAL).

We aim to identify causal relationships among Default-mode and Dorsal Attention networks in resting-state through applying SIA on above-mentioned data. Accordingly, we selected corresponding regions from extracted ROIs as is reported in published literature. The full name, abbreviations, and labels of these ROIs, used in this study, are listed in table (2).

# 3. Results

This section contains the results for implementing SIA on both simulated and real datasets. We will present these results separately in following subsections.

## 3.1. Simulation Results

In this section we will report the results that demonstrate the efficiency of SIA in identifying causal networks in resting brain, based on simulated time-series. To this end, we try to indicate the ability of SIA in discriminating non-zero elements of simulated connectivity matrix $\boldsymbol{A}$, from zero ones, based on the parameter introduced in (38). To put it simply, we redefine the EC estimation in form of a detection problem, and the significance of estimated elements of matrix $\boldsymbol{A}$ cast as the statistics for this detection problem. Therefore, the distribution of these features and the extent of separability of these two distributions represent the ability of SIA in EC detection for each network topology. Figure (5) demonstrates the distribution of parameter $\boldsymbol{S}$, as is defined in (38), for each network topology, for zero and non-zero elements of simulated connectivity matrix. These histograms are plotted for each topology separately, over all corresponding simulated subjects and all the connectivity elements. Since the structure of simulated networks are chosen completely in a random procedure, the complexity of directed graphs constructed based on these network structures are not alike, and consequently, in networks with same number of regions the performance of SIA is not constant. We deliberately conduct the toy examples in this way, in order to investigate the fluctuations in SIA performance for networks with different levels of complexity.

Along with the above-mentioned distributions, for assaying SIA, we also utilized False Positive ratio (FP ratio) and True Positive ratio (TP ratio) for measuring the performance of SIA, defined respectively as the fraction of zero elements of simulated matrix $\boldsymbol{A}$ which have been falsely detected as existent connections, and the fraction of truly detected connections. Figure (6) shows TP and FP ratio for three different threshold values applied to the statistic S, using the null distribution from the phase shuffled surrogate data.

We tried different metrics from graph theory for quantifying the complexity of simulated networks topologies, so that we will be able to better investigate the performance of SIA according to topological characteristics of networks. The metrics we implemented on our networks where previously introduced and interpreted thoroughly in Rubinov and Sporns (2010) and are available via Brain Connectivity Toolbox (https://sites.google.com/a/brain-connectivity toolbox.net/bct/Home). Exploring the correlation of these metrics fluctuations across the topologies and TP ratio demonstrated in figure (6), we found out that the accuracy of EC detection using SIA is highly anti-correlated with the average of *Edge Betweenness Centrality* (EBC) in each network. This metric indicates the fraction of all shortest paths in a directed graph which contain a given edge (Rubinov and Sporns, 2010). The interpretations regarding this metric, and possible conclusions that we can make concerning this observations will be presented in the next section.

In figure (7), the effect of observation noise on detection accuracy is shown. For a specific network topology with constant dynamic noise and HRF across the subjects, we simulated BOLD time-series for different levels of SNR ($SNR: 1 \sim 5\ dB$), and 50 subjects for each SNR level. The TP ratio has been used for indicating this effect. As it was expected, increasing observation noise variance leads to degradation in SIA performance. But, it should be considered that up to a restricted noise level, which is more realistic in fMRI, SIA can retrieve relevant information from noisy observations, and the detection ratio is still above chance level. As explained in section (2.5), we also used simulations for exploring how downsampling can impact SIA performance. Figure (4) displays the network topology on which we have examined the effect of three different TR values. In figure (8) we have plotted the variations in TP and FP ratio for $TR = 1$, 2, 3 seconds. As it is observable in this figure, both the metrics show the robustness of SIA for $TR = 1$, 2 seconds and little degradation in its performance is observed for $TR = 3$ seconds. Based on the topology of figure (4), we also implemented the scenario in which the HRF and neuronal delays oppose each other. For this purpose, we assumed that region 5 drives region 1, as displayed in figure (4), while the HRF in node 1 peaks about 2 seconds before the HRF in node 5. In this case SIA, like methods proposed in (Deshpande et al., 2009; Ryali et al., 2010), is not able to detect the causal interaction correctly.

**Table 1**

Characteristics of simulated networks. In this table the number of regions in each simulated network, the Edge Betweenness Centrality (EBC) measure of them, and standard deviation of elements in vectors $\boldsymbol{b}_m$ can be found.

| Label | # Regions | EBC | HRF std |
|---|---|---|---|
| TOP1_10 | 10 | 0.2 | 0.1 |
| TOP2_10 | 10 | 0.24 | 0.1 |
| TOP3_10 | 10 | 0.24 | 0.1 |
| TOP4_10 | 10 | 0.12 | 0.1 |
| TOP5_10 | 10 | 0.26 | 0.1 |
| TOP6_10 | 10 | 0.07 | 0.1 |
| TOP7_12 | 12 | 0.2 | 0.1 |
| TOP8_12 | 12 | 0.16 | 0.1 |
| TOP9_12 | 12 | 0.26 | 0.1 |
| TOP10_12 | 12 | 0.14 | 0.1 |
| TOP11_12 | 12 | 0.215 | 0.1 |
| TOP12_5 | 5 | 0.2 | 0.1 |
| TOP13_5 | 5 | 0.08 | 0.1 |
| TOP14_5 | 5 | 0.08 | 0.1 |
| TOP15_15 | 15 | 0.08 | 0.1 |
| TOP16_15 | 15 | 0.12 | 0.1 |
| TOP17_15 | 15 | 0.125 | 0.1 |
| TOP18_15 | 15 | 0.09 | 0.1 |
| TOP19_15 | 15 | 0.09 | 0.1 |

According to results reported above, we can see that SIA has this ability to detect EC between brain regions merely based on output BOLD observations. The difference in SIA performance between different topologies is compared to EBC measure for each structure, and we saw that the accuracy of EC detection is highly anti-correlated with EBC values. As we will discuss this observation further in section 4, high EBC values are prevalent in networks with numerous separate modules, which is not the case in brain functional networks according to small world characteristics of these networks. In addition, the simulation results show that thanks to SVD step in SIA, this algorithm is successful in removing the observation noise effect on causality inference up to a specific level. Furthermore, although increasing TR will indirectly lead to reduction of state-space dimension, and decreases the computational complexity of identification algorithm, but simulations demonstrate that due to critical role of temporal resolution in causality analysis, lower TR values result in more accurate EC detection. However, according to simulations, we can at least be sure that for $TR = 2$ seconds, SIA is able to characterize causal networks accurately as well as for $TR = 1$ second, and for lower sampling rates some interactions may be missed, the sensitivity of SIA is still above chance though.

## 3.2. Results of experimental Resting-state fMRI Data:

In this section, we aim to uncover the causal interactions underlying observed anti-correlation between dorsal attention (ATT), and default-mode networks (DMN) during the rest. In other words, we want to investigate which of these two networks place higher than the other in the hierarchy of brain functional networks (He and Raichle, 2009;Carhart-Harris and Friston, 2010). To this end, as explained in previous section, we constructed a network of extracted ROI time-series consists of 16 regions of DMN and ATT networks. For all the ROIs, the time-series related to each subject is concatenated, and one time-series for each ROI is constructed. With this strategy, we have implicitly neglected the inter-subject variability in HRF. Using SIA, we estimated ECs between these regions. Figure (9) demonstrates the causal network constructed from SIA estimated ECs. The threshold for the significance tests were corrected for multiple comparisons by False Discovery Rate (FDR) with $\alpha < 0.05$. In order to answer the question we posed in the beginning of this section, we should specify that the region of which network exert more influences on those of the other one. Therefore, in this graph, we draw the diameter of each node proportionally related to it out-degree (the number of its outgoing connected edges). The label of the nodes used in this figure and correspondent brain regions can be found in table (2). Moreover, two networks are shown in different colors (ATT is white and DMN is red).

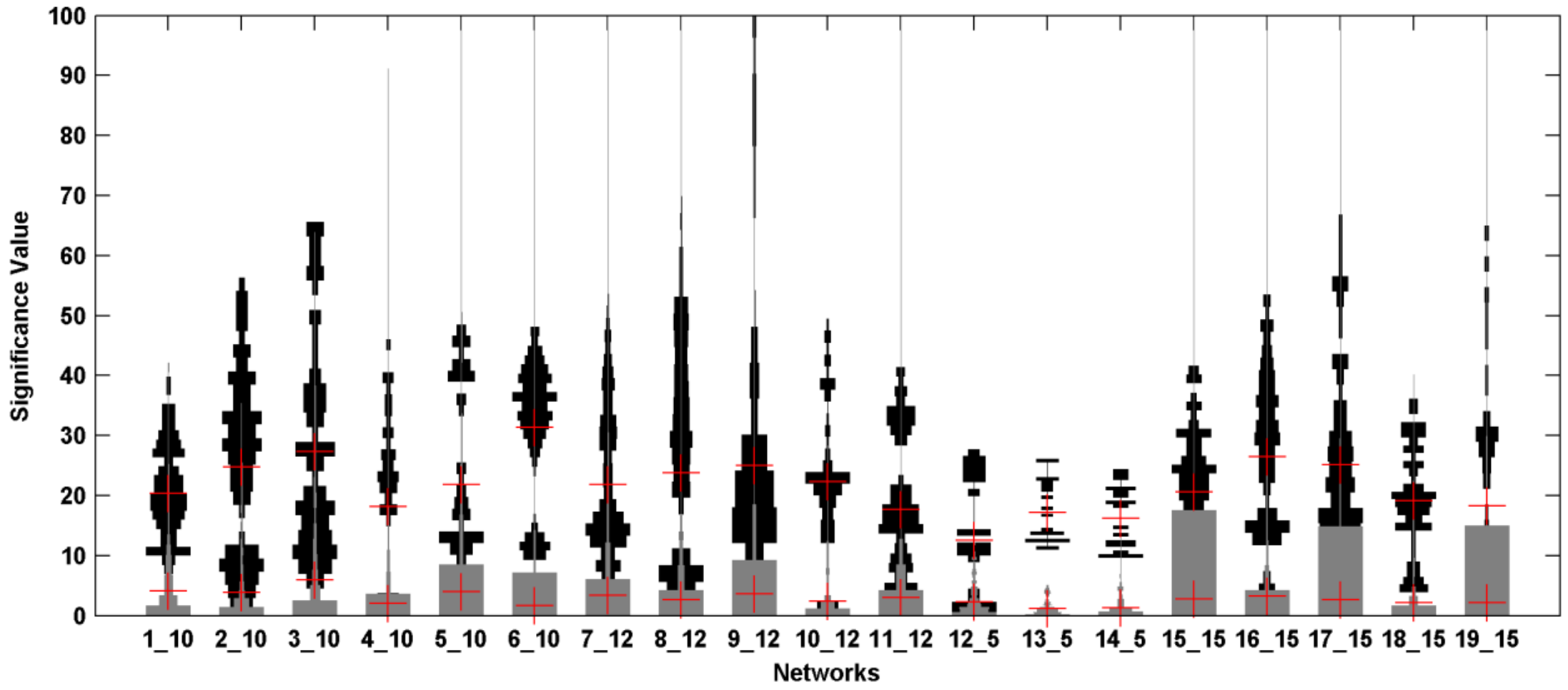

**Fig. 5.** Distribution of significance values (equation (38)) for non-zero (black histograms) and zero (gray histograms) elements of simulated matrix $\boldsymbol{A}$ calculated on estimated matrix $\widehat{\boldsymbol{A}}$. Each column displays these distributions for each network topology listed in table (1). The more non-overlapping these two distributions in each network, the more successful SIA is in estimation of causal interactions in that network. The red crosses indicate the mean value of each distribution. The difference between these mean values for each topology represents the SIA performance in discriminating zero and non-zero elements of matrix $\boldsymbol{A}$.

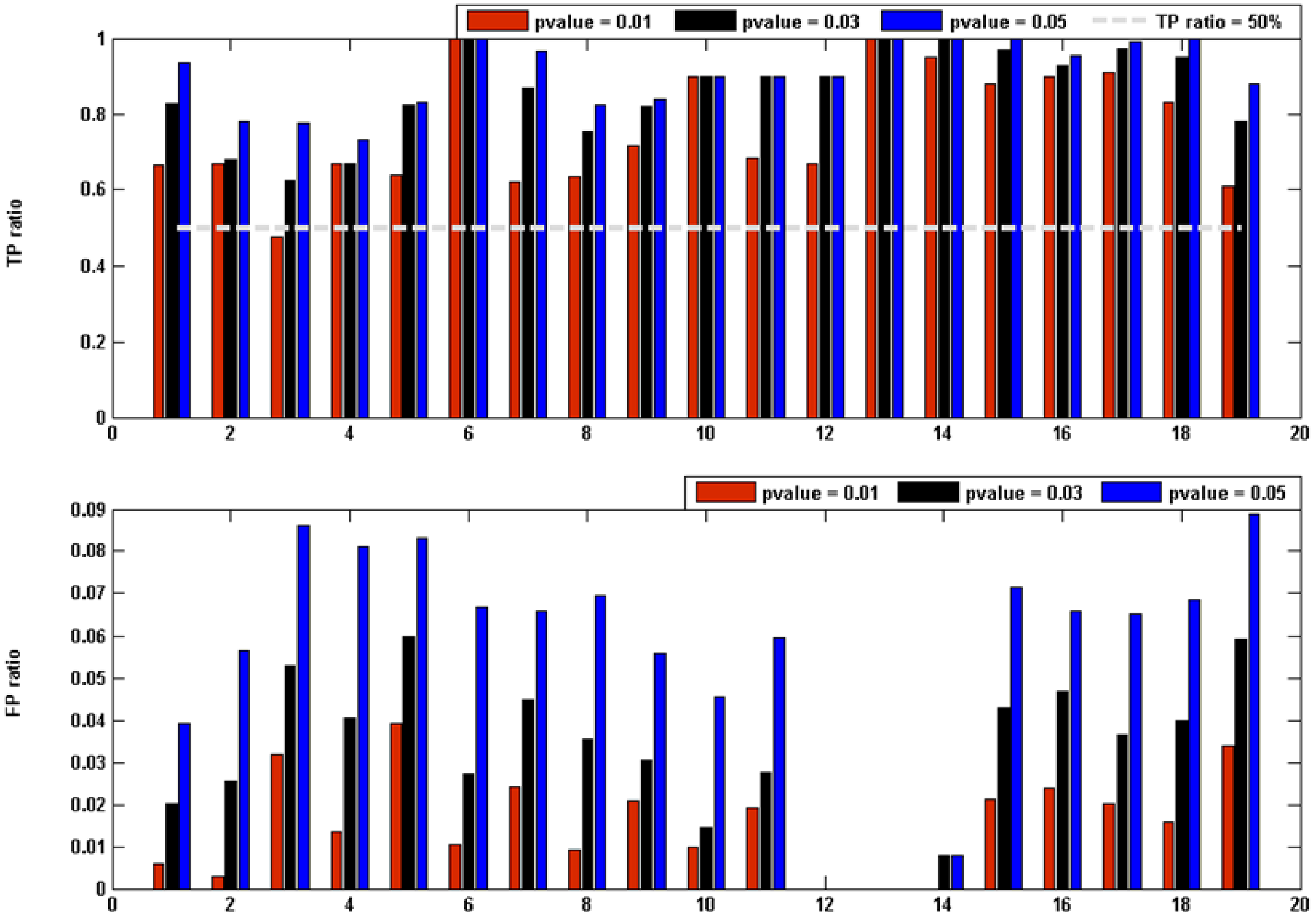

**Fig. 6.** Average of TP and FP ratio in each network across all the corresponding subjects (simulations). In this figure TP ratio (top) and FP ratio (bottom) in each network is averaged across all the subjects. The threshold value is selected for three different $\alpha$ values: $\alpha < 0.05$ (blue), $\alpha < 0.03$ (black), and $\alpha < 0.01$ (red). The white dotted line displays the chance level.

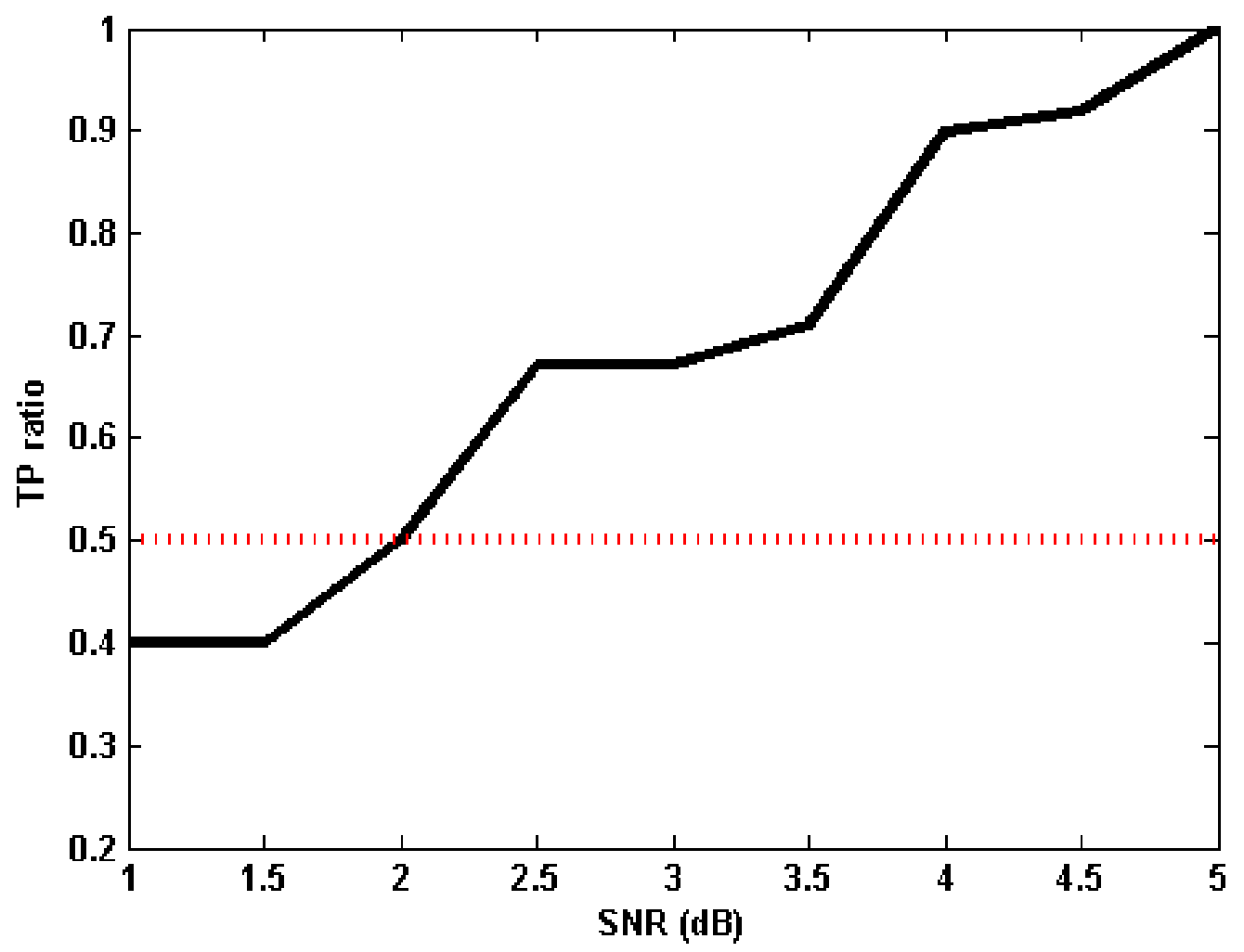

**Fig. 7.** TP ratio for different SNR levels. In order to explore the effects of observation noise on SIA performance, we conduct toy examples with a constant network characteristics and varied SNR level across simulated networks. In this figure, TP ratio is displayed for these simulated networks for different levels of SNR. It is observable that increase in noise standard deviation cause degradation in SIA performance. Red dotted line shows 50 percent sensitivity.

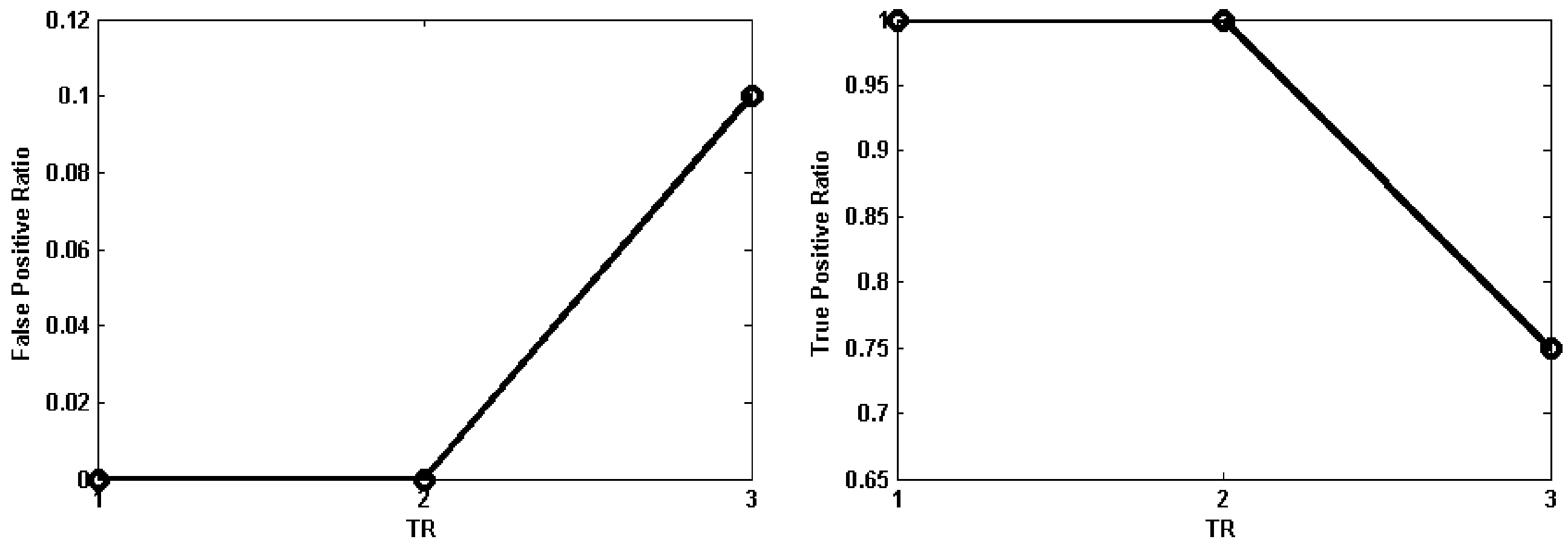

**Fig. 8.** In this figure variation in FP ratio (left side) and TP ratio (right side) for three TR values ($TR = 1, 2, 3\ seconds$) are displayed. Each point is an average over 50 simulations of network in Fig. 4. For $TR = 1$ and $TR = 2$ SIA is totally successful in accurately detecting ECs, while by decreasing the sampling rate SIA accuracy also decreases slightly, but remains in an acceptable level. This test is conducted on the synthetic data simulated based on the network structure depicted in figure (4).

**Table 2**

Default mode and dorsal Attention network. The names of extracted ROIs, the labels used in displayed graph, and their abbreviation is listed in this table.

| Networks | | Regions |
|---|---|---|
| Labels | Abb. | |
| *Default Mode Network* | | |
| 1 | PCC | Posterior Cingulate Cortex |
| 2 | pIPL_L | Left posterior Inferior Parietal Lobule |
| 3 | pIPL_R | Right posterior Inferior Parietal Lobule |
| 4 | vACC | ventral Anterior Cingulate Cortex |
| 5 | dMPFC_L | Left dorsomedial Prefrontal Cortex |
| 6 | dMPFC_R | Right dorsomedial Prefrontal Cortex |
| 7 | DLPFC_L | Left dorsolateral Prefrontal Cortex |
| 8 | DLPFC_R | Right dorsolateral Prefrontal Cortex |
| 9 | PHG_L | Left Parahippocampal Gyrus |
| 10 | PHG_R | Right Parahippocampal Gyrus |
| 11 | ITC_L | Left Inferior Temporal Cortex |
| 12 | ITC_R | Right Inferior Temporal Cortex |
| *Dorsal Attention Network* | | |
| 13 | MT_L | Left Middle Temporal area |
| 14 | MT_R | Right Middle Temporal area |
| 15 | SPL_L | Left Superior Parietal Lobule |
| 16 | SPL_R | Right Superior Parietal Lobule |

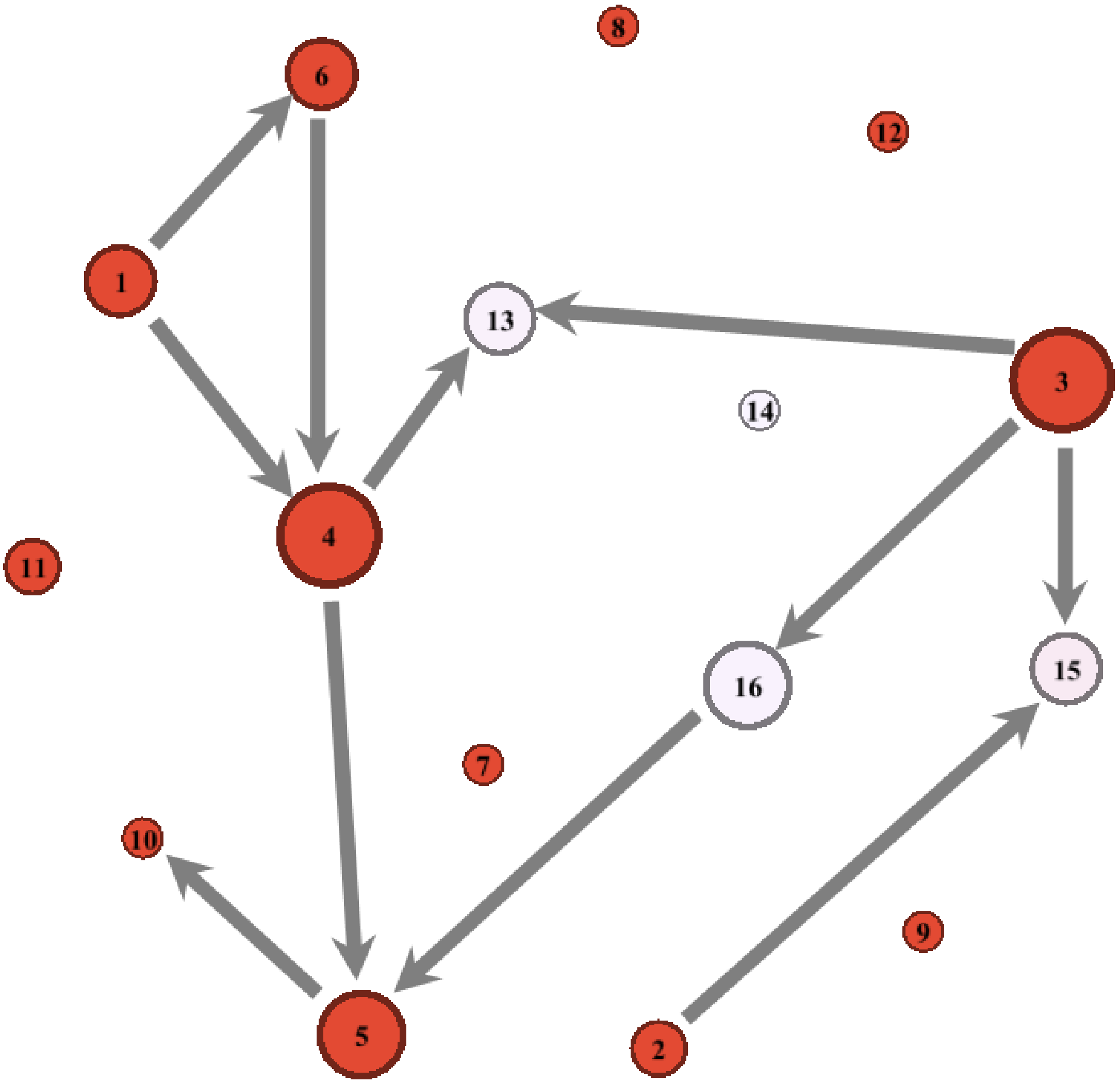

**Fig. 9.** The causal network identified by SIA among default-mode and dorsal attention networks. The label of the nodes and corresponding brain regions are listed in Table (2). In this figure the diameter of the nodes is proportional to their outdegree. For visualizing this graph we used Gephi Graph Visualization and Manipulation software (Bastian et al., 2009).

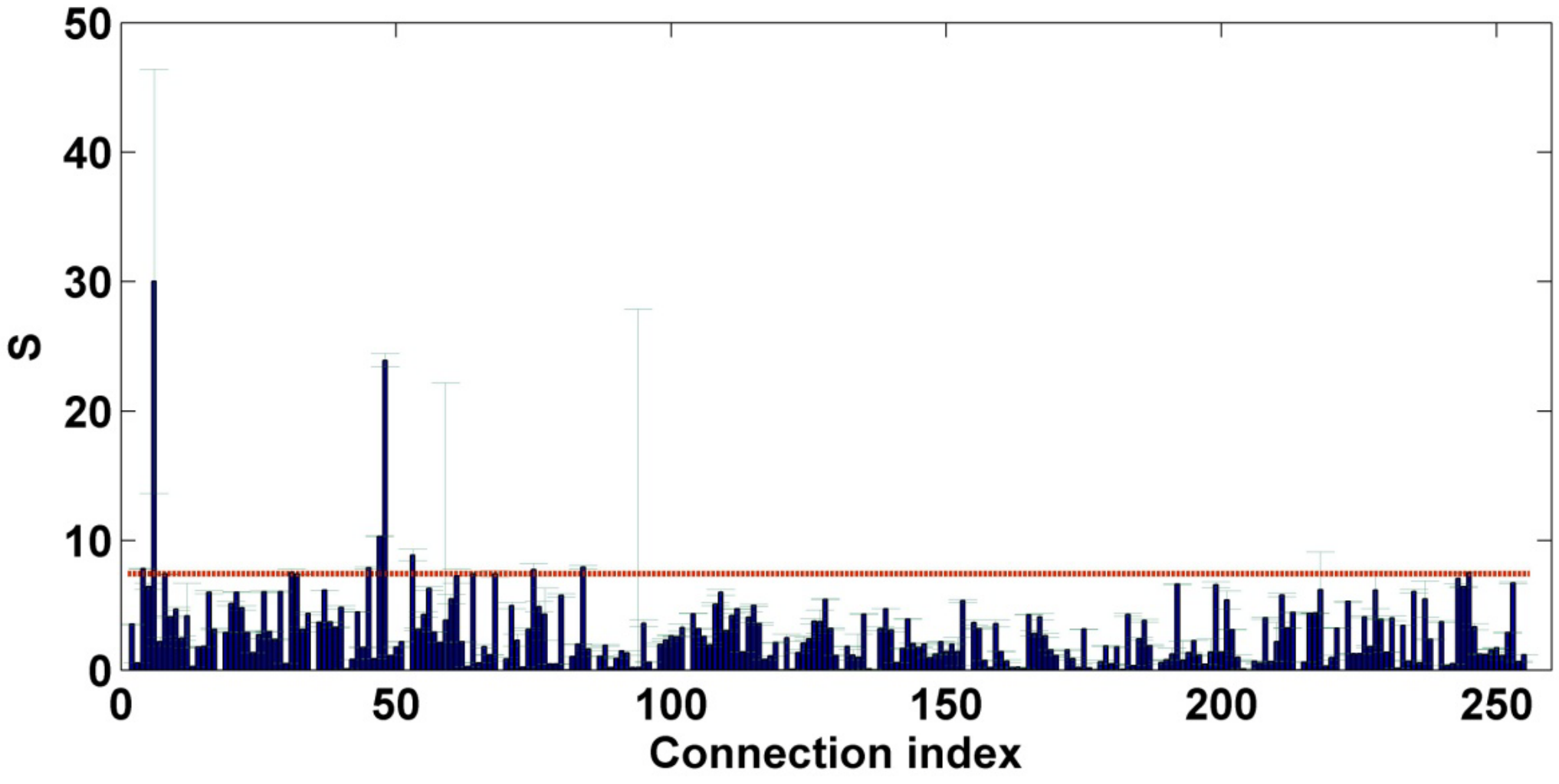

**Fig. 10.** Values of parameter $S$ defined in equation (38) and error bars for estimated elements of matrix $\boldsymbol{A}$ using SIA. The horizontal axis displays the index of elements of matrix $\boldsymbol{A}$, *i.e.* the index of element $a_{ij}$ will be $M \times (j-1) + i$, where $M$ is the number of regions in the model. Blue bars indicate significance of estimated interactions, green lines show error bars of significance values, and red dotted line shows the threshold value (corrected for multiple comparison) corresponding to $\alpha < 0.05$.

Figure (10) indicates the significance of estimated elements of matrix $\boldsymbol{A}$, including their error bars and the threshold (corrected for multiple comparison) corresponding to $\alpha < 0.05$. Above threshold level connections in this figure show the directed edges in figure (9).

Based on characterized causal network in this section, and the graph displayed in figure (9), we conclude that DMN exerts more causal effects on ATT and therefore places higher than ATT in hierarchy of functional brain networks. We will expand more on this observation later in the next section.

# 4 Discussion

In this paper, we developed a state-space based approach for identifying causal networks using rsfMRI. The proposed method acts in a subspace-based system identification framework, and estimate the connectivity matrix through a one –step algebraic process, as discussed in previous sections. Firstly, we applied this algorithm on simulated BOLD time-series, generated from examples of causal networks, in order to assess SIA's abilities in EC detection. In this step, four main aspects of SIA performance were investigated by extensive simulations: ability in discriminating zero and non-zero elements of matrix $\boldsymbol{A}$, TP and FP ratio, and degradations in algorithm performance caused by observation noise and down sampling. Observed fluctuation of SIA performance across different network topologies, inclined us to think that which topological characteristics of a network correlates with degradations in SIA sensitivity. As explained in section (3), the topological metric that best predict these fluctuations in SIA accuracy is EBC, and we saw that increase in EBC is inversely proportional to decrease in TP ratio. The average value of EBC across a network is high, when there are one or more edges in the graph which are shared between many shortest paths. This usually happens when the network topology contains separate communities or modules that are only loosely connected by a few inter-modular edges. In this case, all shortest paths between these communities must run along at least one of these inter-modular edges, so the EBC of these edges, and consequently, the average EBC in the whole network increase (Girvan and Newman, 2002). According to this interpretation, it seems that increase in the modular characteristics of networks topologies is correlated with decrease in SIA performance. Since EBC cannot be confidently calculated for real causal brain network during the rest via any causality analysis, we cannot understand that to what extent SIA is reliable. Nevertheless the small-world characteristic of brain network (Achard and Bullmore, 2007) prevents highly separated communities. The more the structure of network tends to become locally segregated, the less economically will the computations in the brain occur, and it totally contradicts the efficient nature of information integration in brain (Buzsaki, 2006). With this reasoning we expect that EBC remain in a medium level in real networks of resting brain, and consequently, SIA produces acceptable results in real case. In table (1) the characteristics of each simulated network, as well as their EBC values are listed. Standard deviation in regions HRF is considered as the deviations in parameters.

In previous section, we also applied SIA on real resting-state fMRI data, in order to unravel the causal interactions underlying anti-correlation between DMN and ATT. To this end, we used SIA to estimate causal interactions among 16 ROIs of these two networks. Figure (9) displays the result of this implementation. As we explained in that section, we coded the diameter of nodes for determining the out-degree of each node. This can be helpful in obtaining a comprehensive insight about causal interactions underlying resting activities of DMN and ATT. Seven out of eight biggest nodes in this graph belong to DMN, and also three regions of ATT exert no influence on other brain regions. This observation can be interpreted in this way that DMN regions exert more causal influence on ATT. This is an important point in reconstructing the hierarchies of brain functional networks that DMN places higher than ATT in this hierarchy. There were also some speculations about this relationship before in the literature, based on evidences from psychology and resting-state neuroimaging (Carhart-Harris and Friston, 2010) that DMN casts as the highest node in hierarchy of brain networks and can be interpreted as the equivalent of ego-functions in Freudian psychoanalysis. Here we reached a similar conclusion that DMN cause inhibitions in ATT during the rest, rather than vice versa. This conclusion is completely in contradiction with the results reported in Liao et al. (2009) about the order of these two networks in hierarchical structure of resting-state networks. Nevertheless, we believe that this contradiction is mostly caused by spurious inferences in conventional Granger analysis used in Liao et al. (2009). The ability of Granger-like methods in estimating ECs in resting-state has been studied in Smith et al. (2010) through extensive simulations. As we will discuss in next section, the results reported in Smith et al. (2010) and also our simulations in this paper demonstrate poor performance of Granger analysis methods. Therefore, even though Liao et al. (2009) is the first exploratory study for causal inference in resting brain, the results are not reliable due to issues discussed in next section, and results of Smith et al. (2010).

## 4.1 Comparison of SIA and other effective connectivity estimation methods

The approach that we have adopted in section (3) for assessing the algorithm's performance is almost similar to one used in Smith et al. (2010) for comparing the abilities of diverse connectivity analysis methods in fMRI literature. This similarity makes comparisons between SIA and EC analysis methods (discussed in above-mentioned paper) possible. As is best explained in this paper and demonstrated through its extensive simulations, lag-based approaches are quite unsuccessful in directionality retrieval, what is the main purpose of SIA, too. Results that are reported in figure (5) and (6) show the SIA prominence compared to the EC analysis methods discussed in Smith et al. (2010). According to this paper, almost all the lag-based methods discussed in this paper have completely over-lapping zero and non-zero significance distributions; the distributions plotted in figure (5) for SIA. Furthermore, the detection accuracy similar to the metrics we discussed in previous paragraphs and demonstrated in figure (6) is significantly below chance level for these methods. These comparisons clarify that SIA performs much better than previously established EC methods in estimating the directionality of connectivity[1]. We intentionally followed the same approach for presenting results, in order to make them comparable to the comprehensive discussions in

[1] Plots and diagrams presented in Supplementary materials of Smith et al. (2010) contain detailed information about the results presented in this paper. Violin plots, similar to those we presented here in figure (5), can be found therein for networks with different specifications.

Smith et al. (2010). However, we could not apply our developed method on the simulated dataset used in Smith et al. (2010). This restriction mostly comes from the assumptions about dynamic noise in SIA. As mentioned in section (2), even though the distribution of noise is not limited to a specific case, it should be white. However, the paradigm used in Smith et al. (2010) for generating exogenous input to hidden layer violates this important assumption, so we utilized time-series which is simulated based on simple shift in neuronal layer and linear HRF convolution in output layer. Although there are some differences between the models used in our simulations and those carried on in Smith et al. (2010), such as linear vs. nonlinear hemodynamic system model and discrete vs. continuous time state-space formulations, as it is mentioned in that paper too, these differences do not cause any deviations in final conclusions[2]. Taking this point into consideration, we can put our results beside those reported in Smith et al. (2010), and compare the statistics. This comparison leads to this conclusion that SIA is significantly more successful in detecting ECs, and the credit of this success must mostly go to state-space formulation. The main difference between SIA and methods explored in above-mentioned paper is considering hemodynamic system effects on causal inference. In connectivity analysis, particularly in studies which are involved with directionality detection, the temporal aggregation caused by hemodynamic response convolution can mislead connectivity analysis methods toward erroneous inferences. This issue can become more crucial in causality inferences, since temporal aggregation is very likely to destroy lag information in brain region's oscillations. However, the methods, like SIA, which are able to simultaneously deconvolve hemodynamic system effects and estimate causal interactions between time-series, somehow circumvent this obstacle, and based on prior assumptions about hemodynamic system, try to carry on inferences in latent neuronal level. According to this point, we expect that SIA be more successful in estimating direction of causal interactions compared to conventional methods discussed in Smith et al. (2010). An important characteristic of SIA, which makes it eligible for this comparison, is its ability in identifying causal networks with large number of brain regions, and this was not achievable with previously established state-space algorithms.

[2] It is mentioned in Smith et al. (2010), page 2, footnote.

For better understanding the limitations which may arise without considering hemodynamic system effects on causality analysis, we also applied conditional Granger Causality Analysis (cGCA) (Liao et al., 2009) on our simulated dataset and TP ratio and FP ratios are reported in figure (11). For this implementation we used Granger Causal Connectivity Analysis toolbox (Seth, 2010). The order of autoregressive models was determined based on Bayesian Information Criterion (BIC) and Akaike Information Criterion (AIC), and the one with more accurate results is reported in figure (11). As it is apparent in this figure, the difference between SIA and cGCA is mostly in FP ratio. That is to say, neglecting hemodynamic system in cGCA misleads this method mostly toward detecting false connections. For further investigation of this observation, we again used the network in figure (4). The factor that we were looking for in this synthetic experiment was the impact of temporal aggregation on degradation in cGCA performance. To this end, we changed the latency of HRF (the interval between rising and settling time) in region 2 between 7 to 10 seconds and plotted the changes in FP ratio for both SIA and cGCA in figure (12). The dotted connections in figure (4.B) display those which were more likely to be falsely detected by cGCA while increasing the latency of HRF in region 2. We can see in this figure that due to built-in estimation of HRF characteristics in SIA process, it is totally robust against temporal aggregation caused by hemodynamic system, and those false detections do not occur in SIA. Taking these notes into account, we can conclude that considering hemodynamic system effects in estimating causal interactions in SIA lead to more accurate EC detection, compared to temporal precedence-based methods, *e.g.* cGCA, which do not account for these kind of effects in their algorithms.

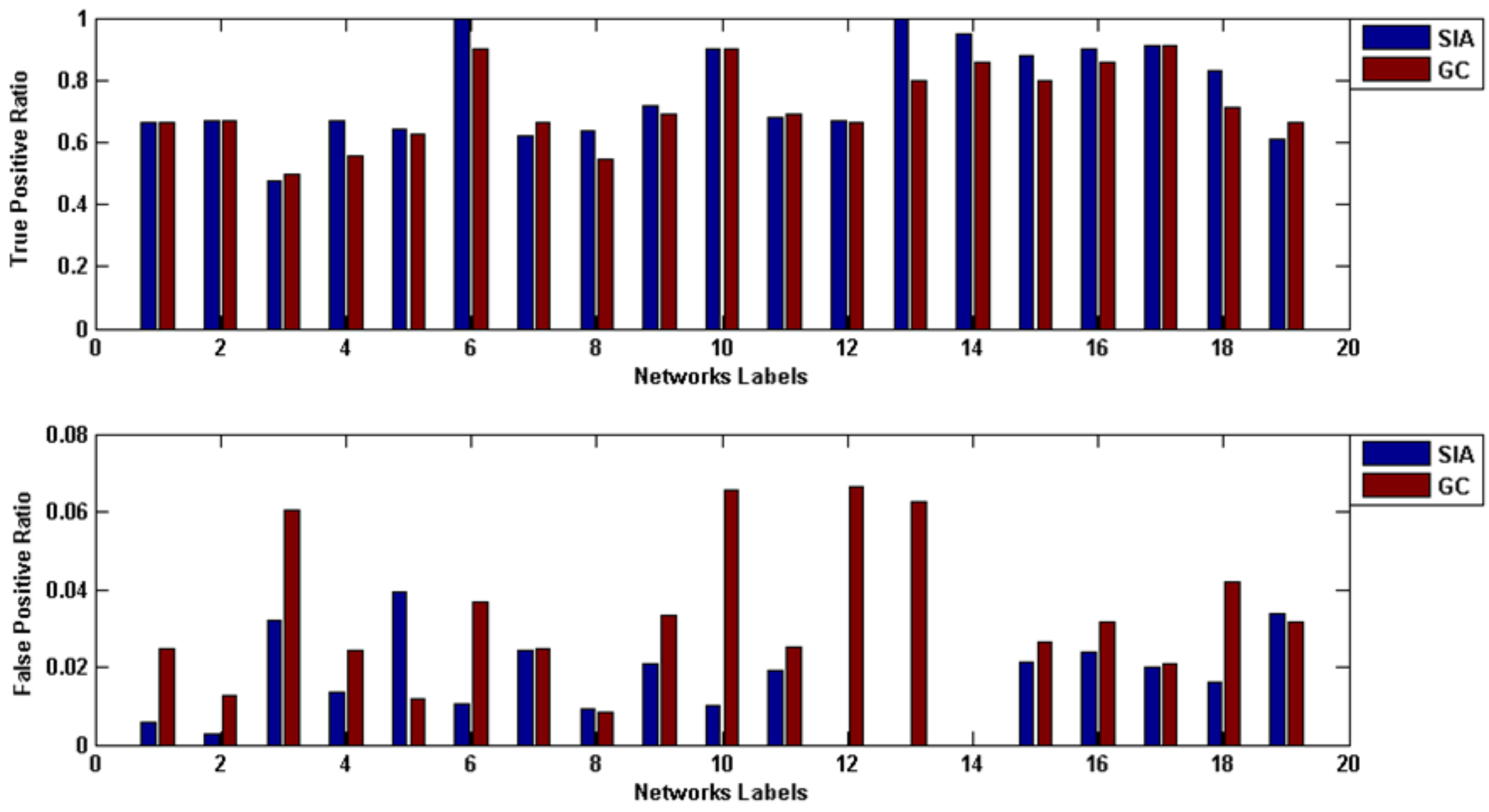

**Fig. 11.** TP and FP ratios for Granger causality analysis (GC) along with SIA on simulated networks in table (1) for threshold of $\alpha < 0.01$. SIA accuracy outperforms cGCA in most of the networks, particularly from FP point of view.

Another method which we tested on our simulated datasets is Expectation-Maximization algorithm (EM) which has been previously proposed for EC estimation based on the state-space model we used in this paper (Ryali et al., 2010; Smith et al., 2009). The implementation has been done according to formulation presented in Smith et al. (2009). It is applied on the simulated networks in table (1) for identifying causal interactions. For all the cases, after 2000 iterations, the algorithm diverges and no acceptable result was obtained. Repeating the procedure for each network and initializing from different points for parameters and state variables do not result in convergence of the iterative algorithm. We examined this process on similar networks, but containing deterministic external stimuli, as well. In these cases, after 15-20 iterations in average, the algorithm converges to optimal solution and results were comparable to those reported in Ryali et al. (2010). However, estimating EC using task-related activity of brain is not the purpose of this study, so we will not expand on the results here. Based on this experiment, we conclude that the role of information conveyed by external stimuli in is too important to be neglected in EM identification of the state-space used in this paper. In absence of this information, the role of initial point become more critical and inappropriate initialization, which cannot be readily handled in EC estimation, consequently leads to divergence of EM algorithm.

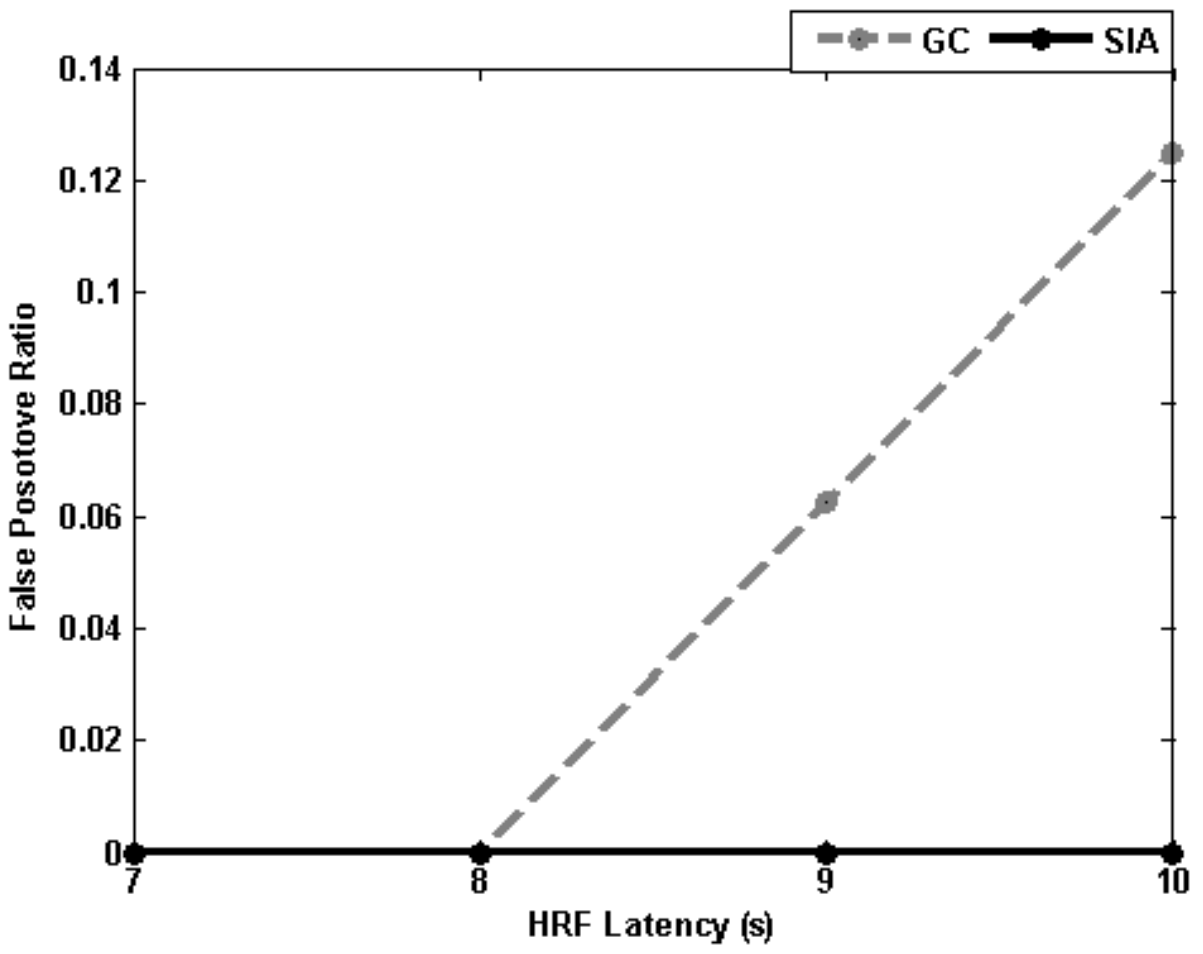

**Fig. 12.** Effect of HRF latency (interval between rising and settling time) on FP ratio in Granger causality analysis (GC) and SIA. Since SIA has a built-in process for estimating characteristics of hemodynamic system in each region, it is more robust than cGCA in dealing with temporal aggregation caused by hemodynamic convolution. This test has been done on simulated data generated based on network structure in figure (4.A). Figure (4.B) shows the connections which are likely to be misclassified as existent connectivity while increasing the HRF latency.

It is worth noting that SIA and its underlying algorithm is adjusted for estimating causal interactions using resting-state brain activity, and does not have any claim regarding EC estimation in task studies. Subspace-based algorithm has been also proposed in system identification literature for linear state-space models with deterministic exogenous inputs (Moonen et al., 1990) which is significantly different in underlying computations from the algorithm proposed for stochastic state-space models. We have not adapted this algorithm for estimating EC using fMRI task studies, and it will be examined in future studies.

## 4.2 Effects of observation noise and down sampling on SIA performance

Another factor that we studied in section (3), was the effect of observation noise on SIA detection accuracy. As displayed in figure (7), SIA is able to cope with observation noise up to a specific level, and any further increase in noise above that level, leads to significant downfall in SIA performance. This can be interpreted by revisiting the SVD step in SIA. In this step we neglect insignificant modes of matrix $\boldsymbol{P}_i$ in subspace reconstruction, and assuredly, low level observation noise can be canceled in this stage through singular value decomposition. Nevertheless, it is applicable until noise deviations are incomparable to signal level, and for higher noise levels, some portion of informative modes diminish among noisy ones, and consequently some lag information will be lost. Therefore, SVD step in SIA is able to implicitly eliminate observation noise and circumvent its effect on causal inference until the noisy modes creep among dominant modes. This happens when the standard deviation of observation noise increases and the informative BOLD time-series is not computationally discernible from noisy signal.

Additionally, we examined the influence of different sampling rates on EC estimation using SIA. The results are displayed in figure (8). As is apparent in this figure, SIA is elegantly successful in revealing neuronal causal interactions from fMRI observations even for $TR = 2$, as well as $TR = 1$. For higher TR values, both sensitivity (TP ratio) and specificity (FP ratio) decreases slightly, but still remain in an acceptable level. As mentioned in section (3), increasing the TR value, will implicitly decrease the dimension of embedded state-space and lead to more economical computations. However, temporal resolution has a crucial role in characterizing causal interactions, and using higher TR values in data acquisition will undoubtedly cause erroneous inferences in causality analysis.

## 4.3 Method limitations and future directions

The approach we have described may have an important role in identifying effective connectivity in the context of variations in the hemodynamic response function using an exploratory approach. Furthermore, we have established a degree of face validity using simulations and real data. In what follows, we contextualize the approach with a deliberate focus on its shortcomings to highlight the work that needs to be done.

First, although we have properly accounted for hemodynamic variability through explicit modeling of latent neuronal states; our discrete time formulation means that our approach is subject to the same criticism as other approaches based upon discrete time auto-regression models. Put simply, the effective connectivity in the $\boldsymbol{A}$ matrix is not the underlying effective connectivity $\boldsymbol{E}$ that one would have obtained using a continuous time formulation. As described in Valdes-Sosa et al. (in press) the two are related through the matrix exponential; where $A = exp(\boldsymbol{E}.TR)$. This means that it is possible for us to infer an element of $\boldsymbol{A}$ is significantly non-zero in the absence of the actual connection. Unfortunately, there is unlikely to be a principled solution to this because our data-led approach (that replies upon subspace identification) cannot accommodate a generative model in continuous time, such as that employed by dynamic causal modeling. However, provided one bears in mind that the effective connectivity may be mediated polysynaptically, this may not be a fundamental problem.

Second, we have to assume that the random fluctuations in our stochastic differential equation that models neuronal activity are Markovian. In other words, they correspond to a Weiner process. This assumption is not plausible in the context of neuronal dynamics, where the endogenous fluctuations are themselves generated by non-Markovian neuronal processes. The extent to which this is a problem could be evaluated in simulations, where we deliberately introduce temporal or serial correlations in the innovations and examine the robustness of SIA to violation of the Markovian assumptions.

Third, we have provided a somewhat heuristic approach to inference on effective connectivity. Crucially, our approach is based upon a null distribution (provided by surrogate data) under the null hypothesis that all effective connectivity between regions is zero. Clearly, this is not an ideal null model if one wants to ask whether a particular connection exists or not. This is because we have assumed that all the other connections are also absent. The usual approach to inference on a single connection is to compare optimized models with and without that connection. However, this requires a particular constraint on the $\boldsymbol{A}$ matrices, which, at present, we have not yet implemented. It is possible that the optimization scheme described in the appendix could be used to place specific constraints on one or a subset of connections; thereby providing the opportunity for formal model comparison and inference about connection strengths of a more rigorous sort. This is the approach taken in dynamic causal modeling and granger causality, where models with and without connections are compared in terms of their Bayesian model evidence.

Fourthly, unlike current approaches to network discovery based on stochastic DCM (Friston et al., 2011), we cannot accommodate nonlinearities in the neuronal model or hemodynamic response function. However, versions of subspace algorithm have been discussed recently in system identification literature for bilinear models as well as more general time-varying models for describing latent layer (Verdult and Verhaegen, 2002; Favoreel, 1999; Chen and Maciejowski, 2000) which can be adapted for estimating EC in discrete state-space models. Although bilinear and more general time-varying models may be unnecessary given the focus on resting-state data, these more complex models are undoubtedly more flexible in modeling the complex dynamics of neuronal interactions. In DCM framework, bilinear connectivity explains the time-varying modulatory influences of external inputs in task studies, while information about modulatory effects in resting-state studies is not accessible. Hence, in resting-state we may consider modulation only caused by stochastic input in bilinear model. Nevertheless, it is not guaranteed that increasing the complexity of the model would certainly lead to more accurate estimation of EC in brain networks, given the low temporal resolution of fMRI data. This is a subtle issue which needs to be carefully examined through generalizability tests, *e.g.* Bayesian comparison test used in conventional DCM or train/test split method.

Finally, although our scheme is non-iterative from the point of view of the subspace identification, it does require iterative optimization to apply the constraints (as described in the appendix). In this sense, it is formally similar to model inversion of the sort used in Bayesian model inversion using filtering or smoothing.

In principal, we can address many of the above issues in terms of robustness to violations of assumptions, using comparative evaluations with network discovery procedures based upon dynamic causal modeling. Dynamic causal models based upon stochastic differential equations of the form in Equation 1 are now available and may provide a useful reference for our subspace based approach. This is potentially important because the computational burden for stochastic dynamic causal models may become prohibitively large when dealing with large numbers of regions in exploratory causality analysis (Lohmann et al., in press). In contrast, our approach can deal gracefully with large networks or graphs.

# 5 Conclusions

In this paper we proposed a novel method for specifying causal interactions in resting brain based on subspace-based identification approach. We examined SIA's abilities in identifying high dimensional causal networks of the brain during the rest. Using fMRI time-series of simulated networks, we investigated the effect of important factors, such as number of brain regions, topological complexity, observation noise, and downsampling on the performance of method. These simulations and the statistical analysis were formed in a manner similar to Smith et al. (2010), so that comparison between SIA and the causality analysis methods considered in this paper became possible. Furthermore we utilized SIA to estimate effective connectivity among brain regions of dorsal attention and default mode network in resting-state, using fMRI data. This study reveals that DMN is at higher order in hierarchy of brain functional networks than ATT.

Although SIA indicates acceptable results in detection of ECs, more complicated computational models of brain dynamics during the rest (Deco et al., 2009) should be used for describing spontaneous interactions in neuronal system, and accordingly, more informative quantitative analysis can take place based on these models. This important issue along with identifiability of more complex neuronal models, using fMRI data, should be addressed in future works.

## Appendix A.

As we explained in (2.3), we need to estimate the optimal linear transformation for transforming identified state-space model (using subspace method) to our desired state-space realization described in equations (6)-(15). For this purpose, here we present a solution through a numerical optimization problem. According to equations (16) and (17):

$$\widetilde{\boldsymbol{A}} = \boldsymbol{T}\widehat{\boldsymbol{A}}\boldsymbol{T}^{-1} \qquad (A.1)$$

$$\boldsymbol{C}\boldsymbol{T} = \boldsymbol{B}\widetilde{\boldsymbol{\Phi}} \qquad (A.2)$$

Based on equation (9), we have defined a specific structure for matrix $\widetilde{\boldsymbol{A}}$. Therefore the first role of linear transformation $\boldsymbol{T}$ is to modify the estimated matrix $\widehat{\boldsymbol{A}}$ to the structure defined in (9). In order to partition product $\boldsymbol{T}\widehat{\boldsymbol{A}}\boldsymbol{T}^{-1}$ into subsections of the desired structure, we define the linear transformation and its inverse according to the following equations:

$$\boldsymbol{T} = \begin{pmatrix} \boldsymbol{U}_1 \\ \boldsymbol{U}_2 \end{pmatrix} \qquad (A.3)$$

$$\boldsymbol{T}^{-1} = (\boldsymbol{V}_1 \quad \boldsymbol{V}_2). \qquad (A.4)$$

Putting these matrices in (A.1) and (A.2) gives:

$$\widetilde{\boldsymbol{A}} = \begin{pmatrix} \boldsymbol{U}_1\widehat{\boldsymbol{A}}\boldsymbol{V}_1 & \boldsymbol{U}_1\widehat{\boldsymbol{A}}\boldsymbol{V}_2 \\ \boldsymbol{U}_2\widehat{\boldsymbol{A}}\boldsymbol{V}_1 & \boldsymbol{U}_2\widehat{\boldsymbol{A}}\boldsymbol{V}_2 \end{pmatrix} \qquad (A.5)$$

$$\widehat{\boldsymbol{C}}\begin{pmatrix}\boldsymbol{U}_1\\ \boldsymbol{U}_2\end{pmatrix}=\boldsymbol{B}\widetilde{\boldsymbol{\Phi}}. \qquad (A.6)$$

Combination of (A.5) and (9), leads to following equation:

$$\begin{pmatrix}\boldsymbol{U}_1\widehat{\boldsymbol{A}}\boldsymbol{V}_1 & \boldsymbol{U}_1\widehat{\boldsymbol{A}}\boldsymbol{V}_2\\ \boldsymbol{U}_2\widehat{\boldsymbol{A}}\boldsymbol{V}_1 & \boldsymbol{U}_2\widehat{\boldsymbol{A}}\boldsymbol{V}_2\end{pmatrix}=\begin{pmatrix}\boldsymbol{A} & \boldsymbol{0}_{M\times M(L-1)}\\ \boldsymbol{J} & \boldsymbol{K}\end{pmatrix} \qquad (A.7)$$

in which $\boldsymbol{J}$ and $\boldsymbol{K}$ are constant matrices. They can be defined from equation (9) in this form:

$$(\boldsymbol{J} \quad \boldsymbol{K})=(\boldsymbol{I}_{M(L-1)} \quad \boldsymbol{0}_{M(L-1)\times M}). \qquad (A.8)$$

Keeping with above equations,

$\boldsymbol{U}_1$ is a $M\times ML$ matrix, $\boldsymbol{U}_2$ is $M(L-1)\times ML$, $\boldsymbol{V}_1$ is $ML\times M$ , and $\boldsymbol{V}_2$ is $ML\times M(L-1)$ and consequently:

$$\begin{pmatrix}\boldsymbol{U}_1\,\boldsymbol{V}_1 & \boldsymbol{U}_1\,\boldsymbol{V}_2\\ \boldsymbol{U}_2\,\boldsymbol{V}_1 & \boldsymbol{U}_2\,\boldsymbol{V}_2\end{pmatrix}=\begin{pmatrix}\boldsymbol{I}_M & \boldsymbol{0}_{M\times M(L-1)}\\ \boldsymbol{0}_{M(L-1)\times M} & \boldsymbol{I}_{M(L-1)}\end{pmatrix}. \qquad (A.9)$$

Based on these equations, we formulated the following optimization problem for determining matrix $\boldsymbol{T}$:

$$\min_{\boldsymbol{U},\boldsymbol{B}} \parallel \widehat{\boldsymbol{C}}\begin{pmatrix}\boldsymbol{U}_1\\ \boldsymbol{U}_2\end{pmatrix}-\boldsymbol{B}\widetilde{\boldsymbol{\Phi}} \parallel_{\mathrm{F}}$$

$$subject\ to\quad \begin{cases}\boldsymbol{U}\boldsymbol{V}=I_{ML}\\ \boldsymbol{U}_1\widehat{\boldsymbol{A}}\boldsymbol{V}_2=\boldsymbol{0}_{M\times M(L-1)}\\ \boldsymbol{U}_2\widehat{\boldsymbol{A}}\boldsymbol{V}_1=\boldsymbol{J}\\ \boldsymbol{U}_2\widehat{\boldsymbol{A}}\boldsymbol{V}_2=\boldsymbol{K}\end{cases} \qquad (A.10)$$

This Nonlinear Programming (NLP) has been solved using interior point line search algorithm (Wachter and Biegler, 2006), implemented in Ipopt software (https://projects.coin-or.org/Ipopt).

## Acknowledgement

We are very grateful to Dr. Babak Nadjar Araabi for helpful discussions which lead to first conjecture about the method presented herein. The authors would like to thank to anonymous reviewers for their thoughtful comments. We are thankful to Jason Smith for providing the implementation of EM algorithm and helpful discussions. We also want to acknowledge ADHD-200 project and the Neuro Bureau from the International Neuroimaging Data-sharing Initiative for providing resting-state fMRI data.

## Captions to Figures:

**Fig. 1.** Basis of Hemodynamic system. Top row: canonical HRF, and bottom row: time derivative of canonical HRF. Linear combination of these two vectors in our model construct HRF of each region . The variations in coefficients of this linear combination, model the inter-region variability in HRFs.

**Fig. 2.** Simulated HRFs. Above displayed HRFs are linear combinations of hemodynamic system basis (Canonical HRF and its time derivative) used in simulations. Three different values for vector $\boldsymbol{b}_m$ have been used for modeling above HRFs.

**Fig. 3.** Sample topology network used in one of the simulations. The black and white square displays the connectivity matrix. White and black squares respectively show non-zero and zero elements in matrix $\boldsymbol{A}$. The schematic connectivity pattern corresponding to this connectivity matrix is also depicted in this figure.

**Fig. 4. (A)** This topology is used for two purposes in this study. First, we used this topology for investigating the effect of TR on SIA performance. BOLD signal for all the regions in this network is simulated for $TR = 1, 2, 3\ seconds$, and the accuracy of SIA in the detection of effective connectivity in this network is displayed in figure (8). Second, we used this network for examining the effect of temporal aggregation caused by HRF convolution on SIA and cGCA results. **(B)** In this figure the dotted connections are prone to be falsely detected as existent connections by conditional Granger Causality Analysis (cGCA) while increasing the latency of HRF in region 2

**Fig. 5.** Distribution of significance values (equation (38)) for non-zero (black histograms) and zero (gray histograms) elements of simulated matrix $\boldsymbol{A}$ calculated on estimated matrix $\widehat{\boldsymbol{A}}$. Each column displays these distributions for each network topology listed in table (1). The more non-overlapping these two distributions in each network, the more successful SIA is in estimation of causal interactions in that network. The red crosses indicate the mean value of each distribution. The difference between these mean values for each topology represents the SIA performance in discriminating zero and non-zero elements of matrix $\boldsymbol{A}$.

**Fig. 6.** Average of TP and FP ratio in each network across all the corresponding subjects (simulations). In this figure TP ratio (top) and FP ratio (bottom) in each network is averaged across all the subjects. The threshold value is selected for three different $\alpha$ values: $\alpha < 0.05$ (blue), $\alpha < 0.03$ (black), and $\alpha < 0.01$ (red). The white dotted line displays the chance level.

**Fig. 7.** TP ratio for different SNR levels. In order to explore the effects of observation noise on SIA performance, we conduct toy examples with a constant network characteristics and varied SNR level across simulated networks. In this figure, TP ratio is displayed for these simulated networks for different levels of SNR. It is observable that increase in noise standard deviation cause degradation in SIA performance. Red dotted line shows 50 percent sensitivity.

**Fig. 8.** In this figure variation in FP ratio (left side) and TP ratio (right side) for three TR values ($TR = 1, 2, 3\ seconds$) are displayed. Each point is an average over 50 simulations of network in Fig. 4. For $TR = 1$ and $TR = 2$ SIA is totally successful in accurately detecting ECs, while by decreasing the sampling rate SIA accuracy also decreases slightly, but remains

in an acceptable level. This test is conducted on the synthetic data simulated based on the network structure depicted in figure (4).

**Fig. 9.** The causal network identified by SIA among default-mode and dorsal attention networks. The label of the nodes and corresponding brain regions are listed in Table (2). In this figure the diameter of the nodes is proportional to their outdegree. For visualizing this graph we used Gephi Graph Visualization and Manipulation software (Bastian et al., 2009).

**Fig. 10.** Values of parameter $S$ defined in equation (38) and error bars for estimated elements of matrix $\boldsymbol{A}$ using SIA. The horizontal axis displays the index of elements of matrix $\boldsymbol{A}$, *i.e.* the index of element $a_{ij}$ will be $M \times (j-1) + i$, where $M$ is the number of regions in the model. Blue bars indicate significance of estimated interactions, green lines show error bars of significance values, and red dotted line shows the threshold value (corrected for multiple comparison) corresponding to $\alpha < 0.05$.

**Fig. 11.** TP and FP ratios for Granger causality analysis (GC) along with SIA on simulated networks in table (1) for threshold of $\alpha < 0.01$. SIA accuracy outperforms cGCA in most of the networks, particularly from FP point of view.

**Fig. 12.** Effect of HRF latency (interval between rising and settling time) on FP ratio in Granger causality analysis (GC) and SIA. Since SIA has a built-in process for estimating characteristics of hemodynamic system in each region, it is more robust than cGCA in dealing with temporal aggregation caused by hemodynamic convolution. This test has been done on simulated data generated based on network structure in figure (4.A). Figure (4.B)

shows the connections which are likely to be misclassified as existent connectivity while increasing the HRF latency.

## Caption to Tables:

**Table 1**

Characteristics of simulated networks. In this table the number of regions in each simulated network, the Edge Betweenness Centrality (EBC) measure of them, and standard deviation of elements in vectors $\boldsymbol{b}_m$ can be found.

**Table 2**

Default mode and dorsal Attention network. The names of extracted ROIs, the labels used in displayed graph, and their abbreviation is listed in this table.